\documentclass[10pt,journal,twoside]{IEEEtran}
\usepackage[applemac]{inputenc}

\ifCLASSINFOpdf
\else
\fi
\usepackage[tight,footnotesize]{subfigure}
\usepackage{epsfig}
\usepackage{amsmath}
\usepackage{amsthm}
\usepackage{amsfonts}
\usepackage{amssymb}
\usepackage{hyperref}
\hypersetup{colorlinks,%
citecolor=black,%
filecolor=black,%
linkcolor=black,%
urlcolor=black,%
pdftex} 
\usepackage{graphicx}
\usepackage{graphics}
\usepackage{float}
\usepackage{url}
\usepackage{eurosym}
\usepackage{setspace}
\usepackage{subfigure}
\usepackage{balance}
\usepackage{dsfont}
\usepackage{amsmath}
\usepackage{bbm}

\newtheorem{theorem}{Theorem}

\def\ontop#1#2{\setbox0\hbox{#2}\copy0\llap{\raise\ht0\hbox{#1}}}

\makeatletter
\renewcommand*\env@matrix[1][\arraystretch]{%
  \edef\arraystretch{#1}%
  \hskip -\arraycolsep
  \let\@ifnextchar\new@ifnextchar
  \array{*\c@MaxMatrixCols c}}
\makeatother

\begin{document}
%
\title{Distributed Diffusion-Based LMS for Node-Specific Adaptive Parameter Estimation}

%
%
%

\author{Jorge~Plata-Chaves,~\IEEEmembership{Member,~IEEE,}
        and~Nikola~Bogdanovi\'{c},~\IEEEmembership{Student Member,~IEEE} 
        and~Kostas~Berberidis,~\IEEEmembership{Senior Member,~IEEE} 
        \thanks{Copyright (c) 2014 IEEE. Personal use of this material is permitted. However, permission to use this material for any other purposes must be obtained from the IEEE by sending a request to pubs-permissions@ieee.org.} 
	\thanks{Jorge Plata-Chaves is with the Department of Electrical Engineering (ESAT-STADIUS), KU Leuven, B-3001 Leuven, Belgium (e-mail: jplata@esat.kuleuven.be as corresponding author).}
	\thanks{Nikola~Bogdanovi\'{c} and Kostas Berberidis are with the Computer Engineering and Informatics Department and RACTI/RU8, University of Patras, Patras 26500 GREECE      
       (e-mails:  \{bogdanovic,berberid\}@ceid.upatras.gr).}
        \thanks{The  work was partially supported by the European SmartEN ITN project (Grant No. 238726) under the Marie Curie ITN FP7 program, by the European HANDiCAMS project (Grant No. 323944) under the Future and Emerging Technologies (FET) programme within the Seventh Framework Programme for Research of the European Commission, and by the University of Patras.}
}

%
%

\markboth{}{PLATA-CHAVES \emph{et al}.: DISTRIBUTED DIFFUSION-BASED LMS FOR NODE-SPECIFIC ADAPTIVE PARAMETER ESTIMATION}
%



\maketitle

\begin{abstract}
A distributed adaptive algorithm is proposed to solve a node-specific parameter estimation problem where nodes are interested in estimating parameters of local interest, parameters of common interest to a subset of nodes and parameters of global interest to the whole network. To address the different node-specific parameter estimation problems, this novel algorithm relies on a diffusion-based implementation of different Least Mean Squares (LMS) algorithms, each associated with the estimation of a specific set of local, common or global parameters. Coupled with the estimation of the different sets of parameters, the implementation of each LMS algorithm is only undertaken by the nodes of the network interested in a specific set of local, common or global parameters. The study of convergence in the mean sense reveals that the proposed algorithm is asymptotically unbiased. Moreover, a spatial-temporal energy conservation relation is provided to evaluate the steady-state performance at each node in the mean-square sense. Finally, the theoretical results and the effectiveness of the proposed technique are validated through computer simulations in the context of cooperative spectrum sensing in Cognitive Radio networks.
\end{abstract}



%
\IEEEpeerreviewmaketitle

\section{Introduction}

\IEEEPARstart{T}{wo} major groups of energy aware and low-complex distributed strategies for estimation over networks have been studied in the literature, i.e., consensus strategies and the algorithms based on incremental or diffusion mode of cooperation. Motivated by the procedure obtained in~\cite{tsitsiklis1986distributed} and~\cite{bertsekasparallel}, the most recent implementations of the consensus strategy (e.g., ~\cite{schizas2009distributed}-\cite{dimakis2010gossip}) allow the cooperating nodes to reach an agreement regarding a vector of parameters of interest in a single time-scale. The second group, which is in the focus of this paper, consists of a single time-scale distributed algorithms that obtain a linear estimator of a vector of parameters by  distributing a specific stochastic gradient method under an incremental or a diffusion mode of cooperation. In the incremental mode (e.g.,\cite{lopes2007incremental}-\cite{li2010distributed}), each node communicates with only one neighbor and the data are processed in a cyclic manner throughout the network. This strategy achieves the centralized-like solution. However, the determination of a cyclic path that covers all nodes of the network is an NP hard problem~\cite{szewczyk2004habitat} and, in addition, cyclic trajectories are more sensitive to node failures and to link failures. Alternatively, better reliability can be achieved at the expense of increased energy consumption in the so-called diffusion mode considered, for instance, in~\cite{lopes2008diffusion}\nocite{cattivelli2010diffusion}-\cite{chouvardas2011}. Under this strategy, each node interacts with a subset of neighboring nodes. As a result, unlike incremental-based strategies, the cooperation is undertaken in a fully ad-hoc fashion. 

In many of the distributed estimation problems, it is considered that the nodes have the same interest. This scenario can be viewed as a special case of a more general problem where the nodes of the network have overlapping but different estimation interests. Some examples of this kind of networks can be found in the context of power system state estimation in smart grids, speech enhancement and active noise control in wireless acoustic networks and cooperative spectrum sensing in Cognitive Radio (CR) networks. Perhaps some of  the first works explicitly considering a network with node-specific estimation interests are~\cite{bertrand2010distributed}-\cite{bertrand2010distributed_smlt}. In these works, for networks with a fully connected and tree topology, Bertrand \emph{et al}. proposed distributed algorithms that allow to estimate node-specific desired signals sharing a common latent signal subspace.

In this paper, we consider the estimation scenarios which can be formulated as Node-Specific Parameter Estimation (NSPE) problems. Within this category, most of the existing works are based on consensus implementations. For instance, the consensus approach presented in~\cite{kekatos2012distributed} is based on optimization techniques that force different nodes to reach an agreement when estimating parameters of common interest. At the same time, the consensus-based technique in~\cite{kekatos2012distributed} allows each node to estimate a vector of parameters that is only of its own interest. In the case of schemes based on a distributed implementation of adaptive filtering techniques, NSPE problems are recently receiving an increasing attention. In~\cite{abdolee2012diffusion}, a diffusion-based scheme is used to solve an NSPE problem where the node-specific estimation interests are expressed as the multiplication of a node-specific matrix of basis functions with a vector of global parameters. Since the matrix of basis functions is known by each node, the problem finally reduces to the estimation of a vector of global interest. In~\cite{chen2012distributed}, the authors use diffusion adaption and scalarization techniques to obtain a Pareto-optimal solution for the the multi-objective cost function that appears in a distributed estimation problem where each node has a different interest.For a network formed by non-overlapping clusters of nodes, each with a different estimation interest, a diffusion-based strategy with an adaptive combination rule is proposed in~\cite{zhao2012clustering}. However, in the proposed strategy the cooperation is finally limited to nodes that have exactly the same objectives. For the same network, Chen \emph{et al}. have recently derived a diffusion-based algorithm with spatial regularization that simultaneously provides biased estimates of the multiple vectors of parameters~\cite{chen2013multitask}. Unlike previous works, the proposed algorithm allows cooperation among neighboring nodes as long as they have numerically similar parameter estimation interests. Additionally, in \cite{chen2014diffusion} the authors analyze the performance of the diffusion-based LMS algorithm derived in~\cite{cattivelli2010diffusion} when it is run in the NSPE setting considered in~\cite{chen2013multitask}.  

As far as the authors are concerned, there are no diffusion-based strategies that provide unbiased solutions of a NSPE problem where the nodes can have overlapping and arbitrarily different estimation interests at the same time. Only in~\cite{bogdanovic2013a}\nocite{platachaves2013a}-\cite{bogdanovic2013aj}, the aforementioned NSPE problem is solved by employing incremental implementations of the Least Mean Squares (LMS) and Recursive Least Squares (RLS) algorithms. Motivated by this fact, we build on our preliminary work~\cite{bogdanovic2014c} in order to design a diffusion-based algorithm that solves a NSPE problem in a network where the nodes can simultaneously be interested in estimating parameters of local, common and/or global interest. In particular, we adopt two peer-to-peer diffusion protocols, Combine-then-Adapt (CTA) and Adapt-then-Combine (ATC), to allow each node to estimate its node-specific vector of parameters in real time under the LMS criterion. Under both CTA and ATC schemes, each node undertakes a local adaptive filtering task where its local observations are fused with an estimate of its parameters of local interest as well as estimates of the parameters of global and common interest, which have been exchanged with its neighbors. As a result, the network is able to adapt in real time to variations of the data statistics related to parameters of local, common and global interest in the network. Moreover, as a detailed performance analysis of the resulting adaptive network shows, the proposed NSPE techniques are asymptotically unbiased in the mean sense. 

The paper is organized as follows. Section~\ref{sec:sec2} mathematically describes the considered NSPE problem. In Section~\ref{sec:sec3} we derive an ATC and CTA diffusion-based techniques to solve the NSPE problem of Section~\ref{sec:sec2} by employing the LMS algorithm. Next, Section~\ref{sec:sec4} is devoted to the theoretical performance analysis of the proposed techniques. Initially, the convergence in the mean sense is analyzed to show that the proposed techniques are asymptotically unbiased. Afterwards, we provide closed-form expressions for the Mean Square Error (MSE) and Mean Square Deviation (MSD) achieved by each node with respect to the estimation of its parameters of local, common and global interest in the steady state. In Section~\ref{sec:sec6}, the theoretical analysis is first verified via generic simulations, and also simulation results are provided in the context of cooperative spectrum sensing in CR networks. Finally, Section~\ref{sec:sec7} summarizes our work and gives a description of the future research lines.

The following notation is used throughout the paper. We use boldface letters for random variables and normal fonts for deterministic quantities. Capital letters refer to matrices and small letters refer to both vectors and scalars. The notation $(\cdot)^H$ and $E\{ \cdot \}$ stand for the Hermitian transposition and the expectation operator, respectively. For a set, e.g., $\mathcal{X}$, the operator $|\cdot|$ stands for the cardinality. If the set $\mathcal{X}$ is ordered, then $\mathcal{X}(j)$ equals the $j$-th element of $\mathcal{X}$. We use the weighted norm notation $\Vert x\Vert^2_{\Sigma}\triangleq x^H\Sigma x$ with a vector $x$ and a Hermitian positive semi-definite matrix $\Sigma \ge 0$. Moreover, $R_{\mathbf{A}} = E \{\mathbf{A}^H \mathbf{A}\}$, $R_{\mathbf{A},\mathbf{B}} = E \{\mathbf{A}^H \mathbf{B}\}$ and $r_{\mathbf{A},\mathbf{b}} = E \{\mathbf{A}^H \mathbf{b}\}$ for any random matrices $\mathbf{A}$, $\mathbf{B}$ and any random vector $\mathbf{b}$. The notation $\mathrm{blockdiag}\{\cdot\}$ denotes a block-diagonal matrix. Finally, $0_{L \times M}$ denotes a $L \times M$ zero matrix, while $1_L$ stands for a $(L \times 1)$ vector of ones. 

\section{Problem statement}
\label{sec:sec2}

Let us consider a connected network consisting of $N$ nodes that are randomly deployed over some geographical region. Nodes that are able to share information with each other are said to be neighbors. The neighborhood of any particular node $k$, including also node $k$, is denoted as $\mathcal{N}_{k}$. Since the network is connected, as shown in Fig.~\ref{fig:fig1}, the neighborhoods are set so that there is a path between any pair of the nodes in the network.

At discrete time $i$, each node $k$ has access to data $ \{d_{k,i}, U_{k,i} \}$, corresponding to time realizations of zero-mean random processes $\{\mathbf{d}_{k,i}, \mathbf{U}_{k,i} \}$, with dimensions $L_k \times 1$ and $L_k \times M_k$, respectively. These data are related to events that take place in the monitored area through the subsequent model
\begin{gather}\label{eps:eq1}
\begin{split}
\mathbf{d}_{k,i} &= \mathbf{U}_{k,i} w_k^o + \mathbf{v}_{k,i} 
\end{split}
\end{gather}
where, for each time instant $i$, 
\begin{itemize}
\item[-] $w_k^o$ equals the deterministic but unknown vector of dimension $M_k$ that gathers all parameters of interest for node $k$,
\item[-] $\mathbf{v}_{k,i}$ denotes the random noise vector with zero mean and covariance matrix $R_{v_k,i}$ of dimensions $L_k \times L_k$,
\item[-] $\mathbf{d}_{k,i}$ and $\mathbf{U}_{k,i}$ are zero-mean random variables with dimensions $L_k \times 1$ and $L_k \times M_k$, respectively.
\end{itemize}
Given the previous observation model, by processing data set $\{d_{k,i}, U_{k,i}\}$ the objective of the network consists in estimating the node-specific vector of parameters $\{w_k\}_{k=1}^{N}$ that minimize the subsequent cost function
\begin{gather}\label{eps:eq2}
\begin{split}
J_{\textrm{glob}}(\{w_k\}_{k=1}^{N}) &=\sum_{k=1}^N E \left \{ \Vert \mathbf{d}_{k,i} - \mathbf{U}_{k,i} w_k \Vert^2 \right \}.
\end{split}
\end{gather}
The vast majority of works dealing with distributed estimation algorithms in the context of adaptive filtering (e.g.,~\cite{lopes2007incremental}\nocite{li2010distributed}-\cite{cattivelli2010diffusion}) considered the case where the nodes' interests are the same, i.e. $w_k^o = w^o$ for all $k \in \{1,2,\ldots,N\}$. However, similarly to~\cite{bogdanovic2013a}\nocite{platachaves2013a}-\cite{bogdanovic2013aj}, the formulation of this paper goes beyond by considering that the node-specific interests are different but overlapping. 

As depicted in Fig.~\ref{fig:fig1}, each node-specific vector $w_k^o$ might consist of a sub-vector $w^o$ of parameters of global interest to the whole network, sub-vectors $\{\varsigma_j^o\}$ of parameters of common interest to subsets of nodes including node $k$, and a sub-vector $\xi_k^o$ of local parameters for node $k$. In particular, the global parameters $w^o$ ($M_g \times 1$) might be related to a phenomenon that can be monitored by all the nodes. In contrast, a set of parameters of common interest $\varsigma_j^o$ ($M_{j_c} \times 1$) might be related to a phenomenon that can be observed by a subset of nodes in the network. The ordered set of indices $k$ associated with the connected subset of nodes interested in $\varsigma_j^o$ is denoted as $\mathcal{C}_j$. For instance, in Fig.~\ref{fig:fig1}, $\mathcal{C}_1=\{1,2,3\}$. Depending on the areas of influence associated with the events of common interest, note that a node might be interested in more than one set of common parameters. As a result, subsets of nodes $\mathcal{C}_j$ and $\mathcal{C}_{j'}$, with $j \neq j'$, might be partially or fully overlapped. For instance, Figure~\ref{fig:fig1} indicates that node $k$ is interested in estimating both vectors of common parameters $\varsigma_j^o$ and $\varsigma_{j-1}^o$, i.e. $\mathcal{C}_{j-1} \cap \mathcal{C}_{j} =\{k\}$. Finally, each vector of local parameters $\xi_k^o$ ($M_{l_k} \times 1$) may represent the influence of some local phenomena that only affects the area monitored by node $k$. In this way, considering a scenario where there are $J$ different subsets of common parameters (see Fig.~\ref{fig:fig1}), the observation model provided in~\eqref{eps:eq1} can be rewritten as
\begin{gather}\label{eps:eq3}
\begin{split}
\mathbf{d}_{k,i} &= \mathbf{U}_{k_g,i} w^o +  \sum_{j \in \mathcal{I}_k } \mathbf{U}_{k_{jc},i} \varsigma_{j}^o  +\mathbf{U}_{k_l,i} \xi_k^o  + \mathbf{v}_{k,i} 
\end{split}
\end{gather}
where, for $k \in \{1,2,\ldots,N\}$, $j \in \{1,2,\ldots,J\}$ and $i \geq 1$, 
\begin{itemize}
\item[-] $\mathcal{I}_k$ equals an ordered set of indices $j$ associated with the vectors $\varsigma_{j}^o$ that are of interest for sensor $k$,
\item[-] $\mathbf{U}_{k_g,i}$,  $\mathbf{U}_{k_{jc},i}$ and $\mathbf{U}_{k_l,i}$ are matrices of dimensions $L_k \times M_{g}$,  $L_k \times M_{j_c}$ and $L_k \times M_{k_l}$ that might be correlated, and  consist of the columns of $\mathbf{U}_{k,i}$ associated with $w^o$,  $ \varsigma_{j}^o$  and $\xi_k^o$, respectively.
\end{itemize}
Thus, according to~\eqref{eps:eq2} and~\eqref{eps:eq3}, our NSPE problem can be restated as minimizing the following cost
\begin{gather}\label{eps:eq4}
\begin{split}
 \sum_{k=1}^N E \left \{ \Vert \mathbf{d}_{k,i} -  \mathbf{U}_{k_g,i} w -  \sum_{j \in \mathcal{I}_k} \mathbf{U}_{k_{jc},i} \varsigma_{j} - \mathbf{U}_{k_l,i} \xi_k \Vert^2 \right \}
\end{split}
\end{gather}
with respect to variables $w, \{\varsigma_{j}\}_{j=1}^J$ and $\{ \xi_k \}_{k=1}^N$.

\section{A solution of the new NSPE problem}
\label{sec:sec3}

In this section, acting as a starting point for the derivation of the distributed algorithms and allowing us to introduce some useful notation, we briefly describe the centralized solution provided in~\cite{bogdanovic2013aj} to the NSPE problem stated in the previous section. Later, via diffusion-based approach we focus on the derivation of distributed algorithms that approximate the centralized solution. For the sake of simplicity and without losing generality, we assume that $M_{l_k}=M_l$, $M_{j_c}=M_c$ and $L_k=L$ for all $k \in \{1,2,\ldots,N\}$ and $j \in \{1,2,\ldots,J\}$. 

\subsection{Centralized solution}\label{subsec:subsec3_1}

An inspection of~\eqref{eps:eq4} reveals that the solution of the considered NSPE problem entails the optimization of a scalar real-valued cost function w.r.t. multiple vector variables, i.e., $\{ w, \{\varsigma_{j}\}_{j=1}^J, \{ \xi_k \}_{k=1}^N \}$. If we gather these variables into the following augmented vector 
\begin{gather}\label{eps:eq7}
\begin{split}
\bar{w}&=\left [w^T \,  \, \varsigma_{1}^T \,  \, \varsigma_{2}^T \cdots  \,  \varsigma_{J}^T \, \, \xi_1^T \, \,  \xi_2^T    \cdots \,  \xi_N^T \right ]^T \quad ( \, \bar{M} \times 1 \, )
\end{split}
\end{gather}
where $\bar{M}=M_g + J\cdot M_c+ N \cdot M_l$, from~\cite{bogdanovic2013aj} we know that our optimization problem can be cast as
\begin{gather}\label{eps:eq5}
\begin{split}
\widehat{\bar{w}}&=\underset{ \tilde{w}  }{\mathrm{argmin}} \left \{ J_{\text{glob}}(\bar{w}) \right \}  \\
&=\underset{ \bar{w}  }{\mathrm{argmin}} \left \{ \sum_{k=1}^N E \left \{ \Vert \mathbf{d}_{k,i} -  \mathbf{\bar{U}}_{k,i} \bar{w} \Vert^2 \right \} \right \}
\end{split}
\end{gather}
where $\mathbf{\bar{U}}_{k,i}$ is defined in~\eqref{eps:eq6} at the top of the following page with $M_{a} =  (k-1) M_l$ and $M_{b}=(N-k) M_l$ and
\begin{figure*}[!ht]
\begin{gather}\label{eps:eq6}
\begin{split}
\mathbf{\bar{U}}_{k,i} = 
\begin{bmatrix}
\mathbf{U}_{k_g,i} & \mathbbm{1}_{\{1 \in \mathcal{I}_k\}} \mathbf{U}_{k_{1c},i} &  \mathbbm{1}_{\{2 \in \mathcal{I}_k\}} \mathbf{U}_{k_{2c},i} &  \cdots  & \mathbbm{1}_{\{J \in \mathcal{I}_k\}} \mathbf{U}_{k_{Jc},i} & 0_{L \times M_{a}} & \mathbf{U}_{k_l,i} & 0_{L \times M_{b}}
\end{bmatrix}
\end{split}
\end{gather}
\rule{\textwidth}{0.1mm}
\end{figure*} 
\begin{gather}\label{eps:eq6b}
\begin{split}
 \mathbbm{1}_{\{ \mathcal{X} \in \mathcal{A} \}} = \left \{
 \begin{array}{ll}
 1 & \mathrm{if} \, \mathcal{X} \subseteq \mathcal{A}, \\
 0 & \mathrm{otherwise.} 
 \end{array}
 \right .
\end{split}
\end{gather}
From~\cite{sayed2011adaptive}, we know that the resulting solutions $\widehat{\bar{w}}$ are optimal if the random processes are $\{\mathbf{d}_{k,i},\mathbf{U}_{k,i}\}$ are jointly wide-sense stationary are given by the normal equations
\begin{gather}\label{eps:eq9}
\begin{split}
\left( \sum_{k=1}^N R_{\bar{U}_{k}} \right ) \cdot \widehat{\bar{w}} =  \sum_{k=1}^N  r_{\bar{U}_{k}d_{k} } .
\end{split}
\end{gather}
Notice that the solution of the previous system of equations requires the transmission of all sensor observations to the fusion center and the inversion of a square matrix whose dimension is proportional to the network size. As a result, for large networks, the centralized solution in~\eqref{eps:eq9} is not scalable with respect to both computational power and communication resources, which motivates the derivation of distributed solutions.

\begin{figure}[t]
  \centering
\centerline{\epsfig{figure=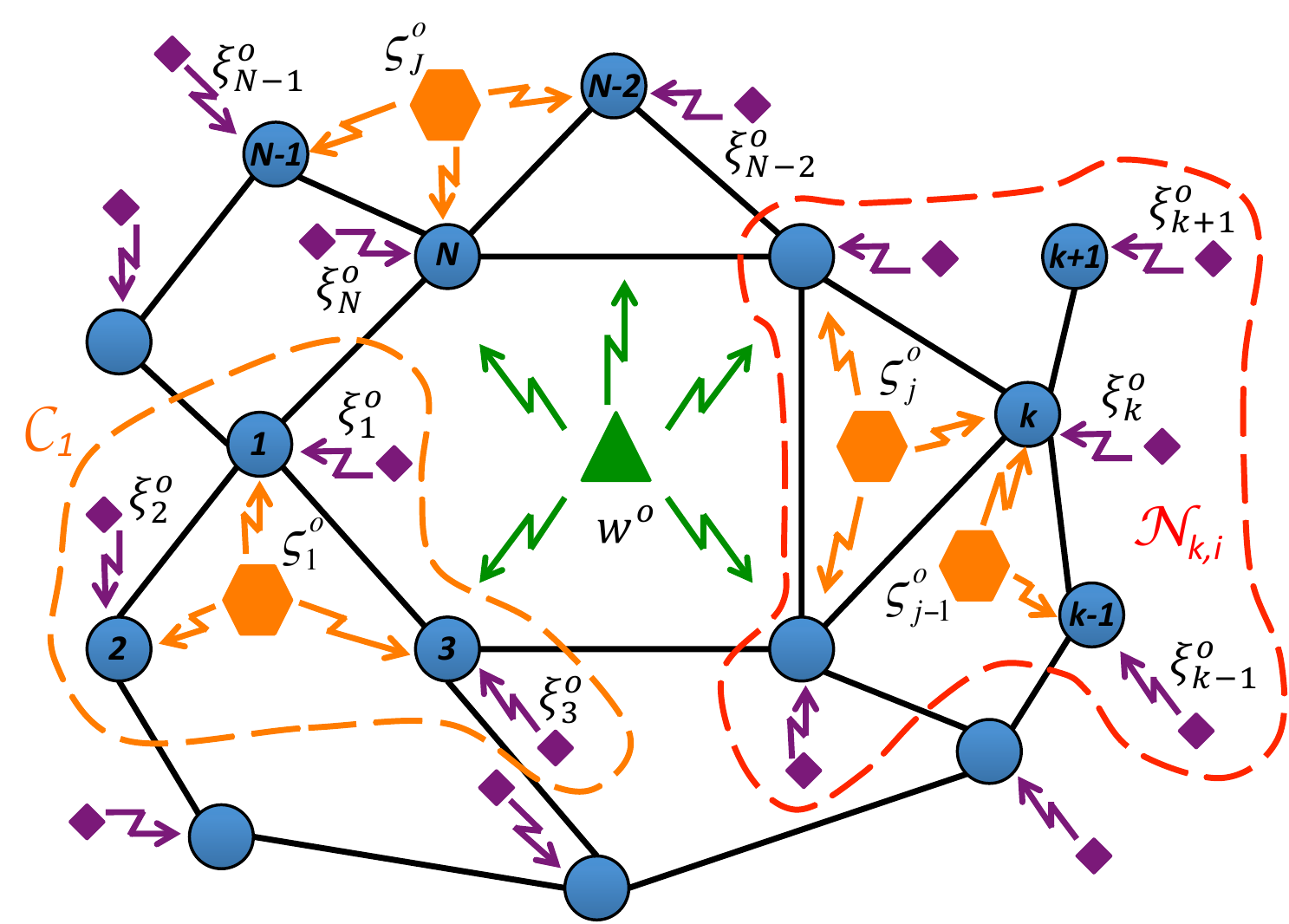,width=8.45cm}}
\caption{A network of $N$ nodes with node-specific parameter estimation interests.}
\label{fig:fig1}
\end{figure}

\subsection{Diffusion-based NSPE solutions}\label{subsec:subsec3_2}

By relying on in-network processing of the data $\{d_{k,i}, U_{k,i}\}$, the incremental-based algorithms proposed in~\cite{bogdanovic2013a} and~\cite{bogdanovic2013aj} converge to the centralized solution in the mean sense with an increase of the energy efficiency and an improved scalability. Attaining more robustness to link or node failures than the incremental strategies, other alternative mode of cooperation to process the data $\{d_{k,i}, U_{k,i}\}$ in a distributed fashion is based on diffusion strategies, e.g., Combine-then-Adapt (CTA) and Adapt-then-Combine (ATC). In the case where the nodes are interested in estimating the same vector of global parameters, the aforementioned strategies are known to well approximate the centralized solution when all nodes want to estimate the same vector of parameters~\cite{cattivelli2010diffusion}. In this work, we extend them so as to be applicable in the NSPE described in Section~\ref{sec:sec2}.

First, let us define $\bar{\psi}_k^{(i)}$ as the local estimate of $\bar{w}^o$ at time instant $i$ and node $k$. Note that $\bar{\psi}_k^{(i)}$ is generally a noisy version of the optimal augmented vector $\bar{w}^o$. By using a diffusion mode of cooperation, at each time instant $i-1$, each node $k$ has access to the set of local estimates of its neighbors, i.e., $\mathcal{N}_{k}$. Thus, node $k$ can fuse its local estimate with the local estimates of its neighbors as follows 
\begin{gather}\label{eps:eq11}
\begin{split}
\bar{\phi}_{k}^{(i-1)}=f_k\left ( \{\bar{\psi}_{\ell}^{(i-1)}\}_{\ell \in \mathcal{N}_{k}} \right )
\end{split}
\end{gather}
where $f_k$ is a local combiner function. In this work, we will focus on linear combiners of the form
\begin{gather}\label{eps:eq12}
\begin{split}
\bar{\phi}_{k}^{(i-1)}=\sum_{\ell \in \mathcal{N}_{k}} \bar{C}_{k,\ell} \, \bar{\psi}_{\ell}^{(i-1)}
\end{split}
\end{gather}
where
\begin{gather}\label{eps:eq13}
\begin{split}
\bar{C}_{k,\ell}  = \mathrm{diag}\{ & c^{ w}_{k,\ell} I_{M_g}, c_{k,\ell}^{\varsigma_1} I_{M_c}, \ldots, c_{k,\ell}^{\varsigma_J} I_{M_c}\\ 
&c_{k,\ell}^{\xi_{1}} I_{M_l}, \ldots,c_{k,\ell}^{\xi_{N}} I_{M_l} \}.
\end{split}
\end{gather} 
In~\eqref{eps:eq13}, $c^{ w}_{k,\ell}$ equals the weight coefficient used by node $k$ when combining the local estimate of the global vector $w^o$ from node $\ell$. Similarly, for $\ell, m \in \{1,\ldots,N\}$, $c_{k,\ell}^{\varsigma_j}$ and $c_{k,\ell}^{\xi_{m}}$ denote the combination coefficients employed by node $k$ when fusing the local estimates of $\varsigma_j^o$ and $\xi_m^o$, from node $\ell$ with its corresponding local estimates, respectively. Since the contribution of each node to the different estimation tasks might be different depending on the statistics of its observations as well as its own estimation interests, note that we allow each node to have different coefficients when combining the local estimates of each vector of global, common or local parameters performed by a neighbor node $\ell$.

To determine the combination coefficients at each node $k$, we can interpret~\eqref{eps:eq12} as a weighted least squares estimate of the augmented vector of parameters $\bar{w}^{o}$ given its local estimate as well as the local estimates at the neighbour nodes~\cite{lopes2008diffusion}. This way, by collecting the local estimates of the augmented vector $\bar{w}^o$ at the neighbour nodes
\begin{gather}\label{eps:eq14}
\begin{split}
\bar{\psi}_{\mathcal{N}_{k}} = \mathrm{col} \left \{ \{ \bar{\psi}_{\ell}^{(i-1)} \}_{\ell \in \mathcal{N}_{k}} \right \}
\end{split}
\end{gather}
and defining 
\begin{gather}\label{eps:eq16}
\begin{split}
B = \mathrm{col} \{I_{\bar{M}}, I_{\bar{M}}, \ldots, I_{\bar{M}} \} \quad \quad (n_{k} \cdot \bar{M} \times \bar{M})
\end{split}
\end{gather}
and
\begin{gather}\label{eps:eq17}
\begin{split}
\bar{C}_k=\mathrm{diag} \{\bar{C}_{k,1}, \bar{C}_{k,2}, \ldots, \bar{C}_{k,n_{k}}\}
\end{split}
\end{gather}
with $n_{k}=|\mathcal{N}_{k}|$, we can formulate the subsequent local weighted least-squares problem
\begin{gather}\label{eps:eq15}
\begin{split}
\underset{ \bar{\phi}_k  }{\mathrm{argmin}} \left \{ \Vert \bar{\psi}_{\mathcal{N}_{k}} -  B \bar{\phi}_k \Vert^2_{\bar{C}_k} \right \},
\end{split}
\end{gather}
whose solution is given by
\begin{gather}\label{eps:eq18}
\begin{split}
\bar{\phi}_k^{(i-1)} = \left [ B^{H} \bar{C}_k B \right ]^{-1} {B}^{H} \bar{C}_k \bar{\psi}_{\mathcal{N}_{k}}.
\end{split}
\end{gather}
In more detail, focusing on the different sub-vectors that form $\bar{\phi}_k^{(i-1)}$, for $k,m \in \{1,2,\ldots,N\}$ and $j \in \{1,2,\ldots,J\}$ the solution provided in~\eqref{eps:eq18} can be rewritten as
\begin{gather}\label{eps:eq19}
\begin{split}
\phi_{k,w}^{(i-1)} = \sum_{\ell \in \mathcal{N}_{k} } \frac{c^{ w}_{k,\ell}}{\sum_{k' \in \mathcal{N}_{k}} c_{k,k'}^{w}} \psi_{\ell,w}^{(i-1)} 
\end{split}
\end{gather}
\begin{gather}\label{eps:eq20}
\begin{split}
\phi_{k, \varsigma_j}^{(i-1)} = \sum_{\ell \in \mathcal{N}_{k} } \frac{ c_{k,\ell}^{\varsigma_j}}{\sum_{k' \in \mathcal{N}_{k}} c_{k,k'}^{\varsigma_j}} \psi_{\ell,\varsigma_j}^{(i-1)}
\end{split}
\end{gather}
and
\begin{gather}\label{eps:eq21}
\begin{split}
\phi_{k,\xi_{m}}^{(i-1)} = \sum_{\ell \in \mathcal{N}_{k} } \frac{c_{k,\ell}^{\xi_{m}}}{\sum_{k' \in \mathcal{N}_{k}} c_{k,k'}^{\xi_{m}}} \psi_{\ell,\xi_{m}}^{(i-1)}
\end{split}
\end{gather}
where $\phi_{k,w}^{(i-1)}$, $\phi_{k, \varsigma_j}^{(i-1)}$ and $\phi_{k,\xi_{m}}^{(i-1)}$ denote the subvectors of combiner $\bar{\phi}_k^{i-1}$ associated with the local estimation of $w^o$, $\varsigma_j^{o}$ and $\xi_{m}$ at node $k$ and time instant $i-1$, respectively. Analogously, $\psi_{k,w}^{(i-1)}$, $\psi_{k, \varsigma_j}^{(i-1)}$ and $\psi_{k,\xi_{m}}^{(i-1)}$ denote the sub-vectors of the local estimate $\bar{\psi}_k^{i-1}$ associated with the local estimation of $w^o$, $\varsigma_j^{o}$ and $\xi_{m}$ at node $j$ and time instant $i-1$, respectively. 

At this point, after a suitable redefinition of the combination coefficients that appear in~\eqref{eps:eq19},~\eqref{eps:eq20} and~\eqref{eps:eq21}, we can now verify that the combination coefficients in~\eqref{eps:eq12} and~\eqref{eps:eq13} have to satisfy
\begin{gather}\label{eps:eq22}
\begin{split}
c^{ w}_{k,\ell} = 0 \textrm{ if } \ell \notin \mathcal{N}_{k}; \quad \sum_{\ell \in \mathcal{N}_{k}} c^{ w}_{k,\ell} = 1
\end{split}
\end{gather}
\begin{gather}\label{eps:eq23}
\begin{split}
c^{\varsigma_j}_{k,\ell} = 0 \textrm{ if } \ell \notin \mathcal{N}_{k}; \quad \sum_{\ell \in \mathcal{N}_{k}} c^{\varsigma_j}_{k,\ell} = 1
\end{split}
\end{gather}
and
\begin{gather}\label{eps:eq24}
\begin{split}
c_{k,\ell}^{\xi_{m}} = 0 \textrm{ if } \ell \notin \mathcal{N}_{k}; \quad \sum_{\ell \in \mathcal{N}_{k}} c_{k,\ell}^{\xi_{m}} = 1
\end{split}
\end{gather}
for $j \in \{1,2,\ldots,J\}$ and $k,m \in \{1,2,\ldots,N\}$.

Next, in order to estimate $\bar{w}^o$ at each node $k$ in an adaptive fashion, the corresponding local aggregate estimate $\bar{\phi}_{k}^{(i-1)}$ is fed into the local LMS-type adaptive algorithm that minimizes the cost associated with node $k$ in~\eqref{eps:eq5}. This way, the resulting diffusion based strategy can be described as
\begin{gather}\label{eps:eq27}
\begin{split}
\left \{
\begin{array}{ll}
\textrm{Combination step:} \\
\bar{\phi}_{k}^{(i-1)}=\sum_{\ell \in \mathcal{N}_{k}} \bar{C}_{k,\ell} \, \bar{\psi}_{\ell}^{(i-1)}  \\
\\
\textrm{Adaptation step:} \\
\bar{\psi}_k^{(i)} = \bar{\phi}_{k}^{(i-1)}  - \mu_k \mathbf{\bar{U}}_{k,i}^H \left[ \mathbf{d}_{k,i} - \mathbf{\bar{U}}_{k,i} \, \bar{\phi}_{k}^{(i-1)} \right ]  \\
\end{array}
\right .
\end{split}
\end{gather}
with $i \geq 1$, $\{\bar{\psi}_\ell^{(0)}\}_{\ell \in \mathcal{N}_k}$ equal to some initial guess, $\bar{C}_{k,\ell}$ defined in~\eqref{eps:eq13} and $\mu_k> 0$  equal to a suitably chosen positive step-size parameter.

Due to the structure of the augmented regressors $\mathbf{\bar{U}}_{k,i}$ defined in~\eqref{eps:eq6}, only $2+|\mathcal{I}_k|$ sub-vectors of $\bar{\psi}_k^{(i)}$ are updated when a specific node $k$ performs the adaptation step at each time instant $i$ (see~\eqref{eps:eq27}). According to~\eqref{eps:eq7} and~\eqref{eps:eq6}, based on $\{\mathbf{d}_{k,i}, \mathbf{U}_{k,i}\}$ and the aggregate estimates $\phi_{k,w}^{(i-1)}$, $\{\phi_{k,\varsigma_j}^{(i-1)}\}_{j \in \mathcal{I}_k}$ and $\phi_{k,\xi_k}^{(i-1)}$, the updated sub-vectors correspond with the local estimates of $w^o$, $\{\varsigma_{k,j}^o\}_{j \in \mathcal{I}_k}$ and $\xi_k^o$ at node $k$ and time $i$, respectively. Therefore, note that each node only updates the sub-vectors that are within its interest, which will be now denoted as $\psi_k^{(i)}$, $\{\varsigma_{k,j}^{(i)}\}_{j \in \mathcal{I}_k}$ and $\xi_k^{(i)}$ for the sake of simplicity. The previous fact allows to set the subsequent equalities in the combination coefficients 
\begin{gather}\label{eps:eq28}
\begin{split}
\left \{
\begin{array}{ll}
c_{k,\ell}^{\xi_{m}} = 0 & \textrm{ if } k \neq \ell \textrm{ or }  k \neq m \\
c_{k,\ell}^{\varsigma_j} = 0 & \textrm{ if } j \notin \mathcal{I}_k \textrm{ or }  j \notin \mathcal{I}_{\ell} 
\end{array}
\right .
\end{split}
\end{gather}
First set of equalities together with~\eqref{eps:eq24} show that $c_{k,k}^{\xi_{k}} = 1$ for each node $k$. Hence, the vector of local parameters $\xi_k^o$ is only estimated by node $k$, which is the only node of the network performing measurements where $\xi_k^o$ is involved. Continuing the analysis of~\eqref{eps:eq28}, from the second set of equalities we can verify that node $k$ only cooperate to estimate the vectors of common parameters that are within its interests, i.e. $\{\varsigma_j^o\}_{j \in \mathcal{I}_k}$. Then, taking into account~\eqref{eps:eq23} we can easily show that 
\begin{gather}\label{eps:eq28a}
\begin{split}
c_{k,\ell}^{\varsigma_j} = 0 \textrm{ if } \ell \notin \mathcal{N}_{k} \cap \mathcal{C}_j; \quad \sum_{\ell \in \mathcal{N}_{k} \cap \mathcal{C}_j} c_{k,\ell}^{\varsigma_j} = 1
\end{split}
\end{gather}
As a result, when a node $k$ estimates a specific vector of common parameters that is within its interest, i.e. $\varsigma_j^o$ with $j \in \mathcal{I}_k$, it will only cooperate with the subset of neighbour nodes $\mathcal{N}_{k} \cap \mathcal{C}_j$, which is composed of the neighbour nodes whose measurements are dependent on $\varsigma_j^o$.

At this point, from~\eqref{eps:eq27} together with~\eqref{eps:eq22}-\eqref{eps:eq24} and~\eqref{eps:eq28}, we can obtain the Combine-then-Adapt (CTA) diffusion-based LMS algorithm summarized below.\\ 
\rule{\linewidth}{0.5mm} \\[-0.5mm]
\textbf{CTA Diffusion-based LMS for NSPE (CTA D-NSPE)}\\[-2mm]
\rule{\linewidth}{0.5mm}
\begin{itemize}
\item Start with some initial guesses $\psi_{k}^{(0)}$, $\{\varsigma_{j}^{(0)}\}_{j \in \mathcal{I}_k}$ 
and $\xi_k^{(0)}$ at each node $k \in \{1,2,\ldots,N\}$ .
\item For the estimation of $w^o$ and any $\varsigma_j^o$, choose $N \times N$ combination matrices $C^{w}$ and $C^{\varsigma_j}$ whose elements in each row $k$, i.e., $\{c_{k,\ell}^{w}\}_{\ell=1}^{N}$ and $\{c_{k,\ell}^{\varsigma_j}\}_{\ell=1}^{N}$, satisfy~\eqref{eps:eq22} and~\eqref{eps:eq28a}, respectively.
\item At each time $i$, for each $k \in \{1,2,\ldots,N\}$, execute
\item[] - Combination step:
\begin{gather}\label{eps:eq30a}
\begin{split}
\phi_{k,w}^{(i-1)} = \sum_{\ell \in \mathcal{N}_{k}} c_{k,\ell}^{w} \, \psi_{\ell}^{(i-1)}
\end{split}
\end{gather}
and
\begin{gather}\label{eps:eq30b}
\begin{split}
\phi_{k,\varsigma_j}^{(i-1)} = \sum_{\ell \in \mathcal{N}_{k} \cap \mathcal{C}_j} c_{k,\ell}^{\varsigma_j} \, \varsigma_{\ell,j}^{(i-1)}
\end{split}
\end{gather}
for each $j \in \mathcal{I}_k$. 
\item[] - Adaptation step:
\begin{gather}\label{eps:eq30c}
\begin{split}
\begin{bmatrix}[1.3] \psi_{k}^{(i)} \\ \varsigma_k^{(i)} \\ 
\xi_k^{(i)} \end{bmatrix}  
= \begin{bmatrix}[1.3]  \phi_{k,w}^{(i-1)} \\  \phi_{k,\varsigma}^{(i-1)} \\
 \xi_k^{(i-1)} \end{bmatrix}  + \mu_k \, U_{k,i}^H \left [ d_{k,i} - U_{k,i}  \begin{bmatrix}[1.3]  \phi_{k,w}^{(i-1)} \\ \phi_{k,\varsigma}^{(i-1)} \\ 
\xi_k^{(i-1)} \end{bmatrix}   \right]
\end{split}
\end{gather}
with $ \varsigma_k^{(i)} = \mathrm{col}\Big \{ \big\{ \varsigma_{k,j}^{(i)} \big \}_{j \in \mathcal{I}_k } \Big \} $ and $ \phi_{k,\varsigma}^{(i)} = \mathrm{col} \Big \{ \big \{ \phi_{k,\varsigma_j}^{(i)} \big \}_{j \in \mathcal{I}_k } \Big \}$.
\end{itemize}
\rule{\linewidth}{0.5mm}\\[-2mm]

Now, let us consider that each node $k$ firstly performs the adaptation step and afterwards, it solves its local weighted least squares problem given in~\eqref{eps:eq15}. Then, by following a derivation that is analogous to the one undertaken for the CTA D-NSPE scheme and that has been omitted for the sake of brevity, we can obtain the Adapt-then-Combine (ATC) diffusion-based LMS algorithm. Basically, as it is summarized in the table shown below, the new NSPE algorithm consists in reversing the order under which the adaptation and combination steps are performed for each node $k$ according to the CTA D-NSPE strategy. \\
\rule{\linewidth}{0.5mm} \\[-0.5mm]
\textbf{ATC Diffusion-based LMS for NSPE (ATC D-NSPE)}\\[-2mm]
\rule{\linewidth}{0.5mm}
\begin{itemize}
\item Start with some initial guesses $\phi_{k,w}^{(0)}$, $\{ \phi_{k,\varsigma_j}^{(0)} \}_{j \in \mathcal{I}_k}$ 
and $\xi_k^{(0)}$ at each node $k \in \{1,2,\ldots,N\}$ .
\item For the estimation of $w^o$ and any $\varsigma_j^o$, choose $N \times N$ combination matrices $C^{w}$ and $C^{\varsigma_j}$ whose elements in each row $k$, i.e., $\{c_{k,\ell}^{w}\}_{\ell=1}^{N}$ and $\{c_{k,\ell}^{\varsigma_j}\}_{\ell=1}^{N}$, satisfy~\eqref{eps:eq22} and~\eqref{eps:eq28a}, respectively.
\item At each time $i$, for each $k \in \{1,2,\ldots,N\}$, execute
\item[] - Adaptation step:
\begin{gather}\label{eps:eq31a}
\begin{split}
\begin{bmatrix}[1.3] \psi_{k}^{(i)} \\ \varsigma_k^{(i)} \\ 
\xi_k^{(i)} \end{bmatrix}  
= \begin{bmatrix}[1.3] \phi_{k,w}^{(i-1)} \\ \phi_{k,\varsigma}^{(i-1)} \\ 
\xi_k^{(i-1)} \end{bmatrix}  + \mu_k \, U_{k,i}^H \left [ d_{k,i} - U_{k,i}  \begin{bmatrix}[1.3]  \phi_{k,w}^{(i-1)} \\ \phi_{k,\varsigma}^{(i-1)} \\ 
\xi_k^{(i-1)} \end{bmatrix}   \right]
\end{split}
\end{gather}
with $ \varsigma_k^{(i)} = \mathrm{col}\Big \{ \big\{ \varsigma_{k,j}^{(i)} \big \}_{j \in \mathcal{I}_k } \Big \} $ and $ \phi_{k,\varsigma}^{(i)} = \mathrm{col} \Big \{ \big \{ \phi_{k,\varsigma_j}^{(i)} \big \}_{j \in \mathcal{I}_k } \Big \}$.
\item[] - Combination step:
\begin{gather}\label{eps:eq31b}
\begin{split}
\phi_{k,w}^{(i)} = \sum_{\ell \in \mathcal{N}_{k}} c_{k,\ell}^{w} \, \psi_{\ell}^{(i)}
\end{split}
\end{gather}
and
\begin{gather}\label{eps:eq31c}
\begin{split}
\phi_{k,\varsigma_j}^{(i)} = \sum_{\ell \in \mathcal{N}_{k} \cap \mathcal{C}_j} c_{k,\ell}^{\varsigma_j} \, \varsigma_{\ell,j}^{(i)}.
\end{split}
\end{gather}
for each $j \in \mathcal{I}_k$. 
\end{itemize}
\rule{\linewidth}{0.5mm}\\[-2mm]

Although the algorithms have been designed for the case where parameters of local, common and global interest coexist, note that the derived algorithms can be simplified straightforwardly to any other NSPE setting. For instance, the derived algorithms can be easily simplified to a setting where there not parameters of global interest or where some of the nodes do not have parameters of local interest. Nevertheless, independently of the considered NSPE setting, we can check that both diffusion-based NSPE algorithms are scalable in terms of computational burden and energy resources. On the one hand, regarding the computational complexity, at each time instant, each node $k$ only needs to update a maximum of $1+ 2 \, (1+|\mathcal{I}_k|)$ vectors whose dimensions are independent of the number of nodes. On the other hand, at each time instant $i$, in both algorithms each node $k$ is required to transmit a maximum of $1+|\mathcal{I}_k|$ vectors, whose dimensions are again independent of the number of nodes.

\section{Performance analysis}\label{sec:sec4}

This section is devoted to the performance analysis of CTA D-NSPE and ATC D-NSPE algorithms proposed in Section III. We start by considering a general recursion that includes both algorithms and that captures the behavior of individual nodes across the network. We then study the convergence in the mean of the general model. Finally, we characterize its mean-square performance in the steady-state in terms of Mean-Square Deviation (MSD) and Excess Mean-Square Error (EMSE).

\subsection{Network-wide recursion}\label{subsec:subsecNetRec}

In this subsection, we derive a general algorithmic form that includes CTA D-NSPE and ATC D-NSPE as special cases. In particular, let us write the first combination step as
\begin{gather}\label{eps:eq_diff_perf_1}
\begin{split}
\phi_{k,w}^{(i-1)} = \sum_{\ell \in \mathcal{N}_{k}} c_{k,\ell}^{w} \, q_{\ell,w}^{(i-1)}
\end{split}
\end{gather}
and
\begin{gather}\label{eps:eq_diff_perf_2}
\begin{split}
\phi_{k,\varsigma_j}^{(i-1)} = \sum_{\ell \in \mathcal{N}_{k} \cap \mathcal{C}_j} c_{k,\ell}^{\varsigma_j} \, q_{\ell,\varsigma_j}^{(i-1)}
\end{split}
\end{gather}
for each $j$ belonging to $\mathcal{I}_k$. Moreover, the adaptation step is expressed in the following
\begin{gather}\label{eps:eq_diff_perf_3}
\begin{split}
\begin{bmatrix}\psi_{k}^{(i)} \\ \ \varsigma_k^{(i)} \\ 
\xi_k^{(i)} \end{bmatrix}  
= \begin{bmatrix}\phi_{k,w}^{(i-1)} \\  \phi_{k,\varsigma}^{(i-1)} \\
 \xi_k^{(i-1)} \end{bmatrix}  + \mu_k \, U_{k,i}^H \left [ d_{k,i} - U_{k,i}  \begin{bmatrix}\phi_{k,w}^{(i-1)} \\ \phi_{k,\varsigma}^{(i-1)} \\ 
\xi_k^{(i-1)} \end{bmatrix}   \right]
\end{split}
\end{gather}
where, with a slight abuse of notation, $\varsigma_k^{(i)} = \mathrm{col} \Big \{ \big \{ \varsigma_{k,j}^{(i)} \big \}_{j \in \mathcal{I}_k } \Big \}$ and $ \phi_{k,\varsigma}^{(i-1)} = \mathrm{col} \Big \{ \big \{ \phi_{k,\varsigma_j}^{(i-1)} \big \}_{j \in \mathcal{I}_k } \Big \}$. The last step of each iteration of the general algorithmic form is the second combination step described as
\begin{gather}\label{eps:eq_diff_perf_4}
\begin{split}
q_{k,w}^{(i)} = \sum_{\ell \in \mathcal{N}_{k}} a_{k,\ell}^{w} \, \psi_{\ell}^{(i)}
\end{split}
\end{gather}
and
\begin{gather}\label{eps:eq_diff_perf_5}
\begin{split}
q_{k,\varsigma_j}^{(i)} = \sum_{\ell \in \mathcal{N}_{k} \cap \mathcal{C}_j} a_{k,\ell}^{\varsigma_j} \, \varsigma_{\ell,j}^{(i)}.
\end{split}
\end{gather}
for each $j$ belonging to $\mathcal{I}_k$. In~\eqref{eps:eq_diff_perf_1}, the non-negative real coefficient $c_{k,\ell}^{w}$ corresponds to the $(k,\ell)$-th entries of the $(N \times N)$ combination matrix $C^{w}$, which satisfies $C^{w}1_{N}=1_{N}$.
Moreover, in~\eqref{eps:eq_diff_perf_2}, the non-negative real coefficient $c_{k,\ell}^{\varsigma_j}$ corresponds to the $(|\mathcal{C}_{j,k}|,|\mathcal{C}_{j,\ell}|)$ entry of a $(|\mathcal{C}_j| \times |\mathcal{C}_j|)$ combination matrix $C^{\varsigma_j}$, which satisfies $C^{\varsigma_j}1_{|\mathcal{C}_j|}=1_{|\mathcal{C}_j|} $
with
\begin{equation}\label{eps:eq_def_cjk}
\mathcal{C}_{j,k} = \{k' \in \mathcal{C}_{j}: k' \leq k \}.
\end{equation}
and $k,\ell \in \mathcal{C}_j$ for any $j \in \{1,2,\ldots,J\}$. Similarly, in~\eqref{eps:eq_diff_perf_4}-\eqref{eps:eq_diff_perf_5} the non-negative real coefficients $\{a_{k,\ell}^{w}\}$ and $\{a_{k,\ell}^{\varsigma_j}\}$ correspond to the $(k,\ell)$-th and the $(|\mathcal{C}_{j,k}|,|\mathcal{C}_{j,\ell}|)$-th entries of the $(N \times N)$ and $(|\mathcal{C}_j| \times |\mathcal{C}_j|)$ combination matrices $A^{w}$ and $A^{\varsigma_j}$, respectively, which satisfy
\begin{displaymath}
A^{w}1_{N}=1_{N}, \qquad A^{\varsigma_j}1_{|\mathcal{C}_j|}=1_{|\mathcal{C}_j|}
\end{displaymath}
for any $j \in \{1,2,\ldots,J\}$.

Also, note that if we set $C^{w}=I_{N}$, $C^{\varsigma_j}=I_{|\mathcal{C}_j|}$ for $j \in \{1,2,\ldots,J\}$, equations~\eqref{eps:eq_diff_perf_1}-\eqref{eps:eq_diff_perf_5} represent ATC D-NSPE. On the other hand, its CTA counterpart corresponds to selecting $A^{w}=I_{N}$, $A^{\varsigma_j}=I_{|\mathcal{C}_j|}$ for $j \in \{1,2,\ldots,J\}$.

Now, let us interpret data as random variables. Associated with the quantities in the general form in~\eqref{eps:eq_diff_perf_1}-\eqref{eps:eq_diff_perf_5}, we define the weight-error vectors, for $k=\{1, \ldots , N\}$ and $j=\{1, \ldots , J\}$, as follows
\begin{gather}\label{eps:eq_diff_perf_1_1}
\begin{split}
\boldsymbol{\tilde{\phi}}_{k,w}^{(i)}= w^o -\boldsymbol{\phi}_{k,w}^{(i)}, \, \boldsymbol{\tilde{p}}_{k,w}^{(i)}= w^o - \boldsymbol{\psi}_{k}^{(i)}, \, \boldsymbol{\tilde{q}}_{k,w}^{(i)}= w^o -\boldsymbol{q}_{k,w}^{(i)}  \\
\boldsymbol{\tilde{\phi}}_{k,\varsigma_j}^{(i)}= \varsigma_j^o - \boldsymbol{\phi}_{k,\varsigma_j}^{(i)}, \, \boldsymbol{\tilde{p}}_{k,\varsigma_j}^{(i)}= \varsigma_j^o - \boldsymbol{\varsigma}_{k,j}^{(i)}, \, \boldsymbol{\tilde{q}}_{k,\varsigma_j}^{(i)}= \varsigma_j^o - \boldsymbol{q}_{k,\varsigma_j}^{(i)}  \,  \,\\
\boldsymbol{\tilde{\phi}}_{k,\xi_k}^{(i)}= \xi_k^o - \boldsymbol{\xi}_k^{(i)}, \, \boldsymbol{\tilde{p}}_{k,\xi_k}^{(i)}= \xi_k^o - \boldsymbol{\xi}_k^{(i)} , \,  \boldsymbol{\tilde{q}}_{k,\xi_k}^{(i)}= \xi_k^o - \boldsymbol{\xi}_k^{(i)}  . \,  \,  \,
\end{split}
\end{gather}

Next, we collect these quantities across all agents into the corresponding $(\sum_{k=1}^N M_k \times 1)$ block vectors, i.e., network weight-error vectors,
\begin{gather}\label{eps:eq_diff_perf_6}
\begin{split}
\boldsymbol{\tilde{\phi}}_{i}= \mathrm{col}\left \{ \left \{\boldsymbol{\tilde{\phi}}_{k,w}^{(i)}, \{\boldsymbol{\tilde{\phi}}_{k,\varsigma_j}^{(i)}\}_{j \in \mathcal{I}_k },\boldsymbol{\tilde{\phi}}_{k,\xi_k}^{(i)}, \right \}_{k=1}^N \right \}
\end{split}
\end{gather}
In the same vein, the network vectors $\boldsymbol{\tilde{p}}_{i}$ and $\boldsymbol{\tilde{q}}_{i}$ are formed, by stacking the corresponding weight-error vectors. For notational convenience, hereafter we use 
\begin{gather*}
\begin{split}
\breve{M}=\sum_{k=1}^N M_k \,.
\end{split}
\end{gather*}
To proceed, let us introduce the diagonal matrix 
\begin{gather}\label{eps:eq_diff_perf_6}
\begin{split}
\mathcal{M}= \mathrm{diag}\{\mu_1 I_{M_1}, \ldots, \mu_N I_{M_N}  \} \qquad (\breve{M} \times \breve{M}),
\end{split}
\end{gather}
the block-diagonal matrix
\begin{gather}\label{eps:eq_diff_perf_7}
\begin{split}
\boldsymbol{\mathcal{D}}_i= \mathrm{diag}\{\mathbf{U}_{1,i}^H \mathbf{U}_{1,i}, \ldots, \mathbf{U}_{N,i}^H \mathbf{U}_{N,i}  \} \qquad (\breve{M} \times \breve{M}),
\end{split}
\end{gather}
and the vector
\begin{gather}\label{eps:eq_diff_perf_8}
\begin{split}
\boldsymbol{\mathcal{V}}_i= \mathrm{col}\{\mathbf{U}_{1,i}^H \mathbf{v}_{1,i}, \ldots, \mathbf{U}_{N,i}^H \mathbf{v}_{N,i}  \} \qquad (\breve{M} \times 1).
\end{split}
\end{gather}

Finally, the network-wide behavior can be characterized by these relations for the block quantities:
\begin{gather}\label{eps:eq_diff_perf_9}
\begin{split}
\boldsymbol{\tilde{\phi}}_{i-1}= \breve{\mathcal{C}} \, \boldsymbol{\tilde{q}}_{i-1}
\end{split}
\end{gather}
\begin{gather}\label{eps:eq_diff_perf_10}
\begin{split}
\boldsymbol{\tilde{p}}_{i}= (I-  \mathcal{M}\boldsymbol{\mathcal{D}}_i)\boldsymbol{\tilde{\phi}}_{i-1} - \mathcal{M} \boldsymbol{\mathcal{V}}_i
\end{split}
\end{gather}
\begin{gather}\label{eps:eq_diff_perf_11}
\begin{split}
\boldsymbol{\tilde{q}}_{i}= \breve{\mathcal{A}} \, \boldsymbol{\tilde{p}}_{i}
\end{split}
\end{gather}
where the structure of the extended weighting matrices $\breve{\mathcal{A}}$ and $\breve{\mathcal{B}}$ is explained in the following subsection.  

Note that equations~\eqref{eps:eq_diff_perf_9}-\eqref{eps:eq_diff_perf_11} can be summarized in the following equivalent form, 
\begin{gather}\label{eps:eq_diff_perf_12}
\begin{split}
\boldsymbol{\tilde{q}}_{i}= \breve{\mathcal{A}} \, (I-  \mathcal{M}\boldsymbol{\mathcal{D}}_i)   \,\breve{\mathcal{C}} \, \boldsymbol{\tilde{q}}_{i-1} - \breve{\mathcal{A}} \,\mathcal{M} \boldsymbol{\mathcal{V}}_i,
\end{split}
\end{gather}
which will be used in the Subsections~\ref{subsec:subsecMeanStability} and~\ref{subsec:subsecSteadyState} to perform the mean and the mean-square steady-state analysis, respectively.

\subsection{Structure of the extended weighting matrices}\label{subsec:subsecWeightMatrix}

The extended weighting matrices $\breve{\mathcal{C}}$ and $\breve{\mathcal{A}}$ have the same form, only the weights are different. Therefore, in order to define them, let us consider, for instance, the  $\breve{\mathcal{A}}$ matrix,  
\begin{gather} \label{eps:eq_EWM_Jorge1}
\begin{split}
\breve{\mathcal{A}} = 
\mathrm{col} \left \{A_1^w , \{A_1^{\varsigma_j}\}_{j \in \mathcal{I}_1 }, A_1^{\xi_1}, \ldots, A_N^w , \{A_N^{\varsigma_j}\}_{j \in \mathcal{I}_N }, A_N^{\xi_N} \right \},
\end{split}
\end{gather}
where the blocks being stacked are defined in~\eqref{eps:eq_EWM_Jorge2}-\eqref{eps:eq_EWM_Jorge5}  on the top of the following page, with $\mathcal{I}_{jk}$ in~\eqref{eps:eq_EWM_Jorge4} defined as $\mathcal{I}_{jk}= \{j' \in \mathcal{I}_{k} : j' <j \}$.

\begin{figure*}[!ht]
\begin{gather}\label{eps:eq_EWM_Jorge2}
\begin{split}
A_k^w = 
\begin{bmatrix}
a_{k,1}^w I_{M_g} & 0_{M_g \times (M_c |\mathcal{I}_1| +M_l)} &  a_{k,2}^w I_{M_g} & 0_{M_g \times (M_c |\mathcal{I}_2| +M_l)} & \ldots & a_{k,N}^w I_{M_g} & 0_{M_g \times (M_c |\mathcal{I}_N| +M_l)}
\end{bmatrix}
\end{split}
\end{gather}
\begin{gather}\label{eps:eq_EWM_Jorge3}
\begin{split}
A_k^{\varsigma_j} = 
\begin{bmatrix}
A_{k1}^{\varsigma_j} & A_{k2}^{\varsigma_j} &   \ldots & A_{kN}^{\varsigma_j}
\end{bmatrix}
\end{split}
\end{gather}
\begin{gather}\label{eps:eq_EWM_Jorge4}
\begin{split}
 A_{k\ell}^{\varsigma_j} = \left \{
 \begin{array}{ll}
 \begin{bmatrix}
0_{M_c \times (M_g + M_c |\mathcal{I}_{j\ell}| )} & a_{k,\ell}^{\varsigma_j}  I_{M_c} &  0_{M_c \times ( M_c [|\mathcal{I}_{\ell}|-|\mathcal{I}_{j\ell}|-1] +M_l)} 
\end{bmatrix} & \mathrm{if} \, \ell \in  \mathcal{C}_j, \\
 0_{M_c \times M_{\ell}} &  \mathrm{if} \, \ell \not\in  \mathcal{C}_j 
 \end{array}
 \right .
\end{split}
\end{gather}
\begin{gather}\label{eps:eq_EWM_Jorge5}
\begin{split}
A_{k}^{\xi_k} = 
\begin{bmatrix}
0_{M_l \times (\sum_{\ell=1}^{k-1} M_{\ell} + M_g + M_c |\mathcal{I}_{k}| )}  & I_{M_l} &   0_{M_l \times (\sum_{\ell=k+1}^N M_{\ell} )}
\end{bmatrix}
\end{split}
\end{gather}
\rule{\textwidth}{0.1mm}
\end{figure*} 

An alternative way to define $\breve{\mathcal{A}}$ is the following relation
\begin{gather}\label{eps:eq_EWM_P1}
\begin{split}
\breve{\mathcal{A}}= \mathcal{P} \, \breve{\mathcal{A}}^{\mathrm{blkd}} \, \mathcal{P}^T
\end{split}
\end{gather}
where the block-diagonal matrix is given by
\begin{gather}\label{eps:eq_EWM_P2}
\begin{split}
\breve{\mathcal{A}}^{\mathrm{blkd}} = \mathrm{blockdiag} \left \{ A^{w}\otimes I_{M_g},  \{A^{\varsigma_j} \otimes I_{M_c} \}_{j =1 }^J,  I_N\otimes I_{M_l} \right \}
\end{split}
\end{gather}
while $\otimes$ stands for the Kronecker product, and $\mathcal{P}$ is the $\breve{M} \times \breve{M}$ permutation matrix that stacks appropriately chosen $1 \times \breve{M}$ row basis vectors. In particular, a basis vector $e_k$ has the unity at the $k^{th}$ position and zeros elsewhere. For more details how the $\mathcal{P}$ matrix is specified, see Appendix~\ref{ap:appendix1}.

\subsection{Data assumptions}\label{subsec:subsecDataAssumptions}

To proceed, we state the following independence assumptions on the data:
\begin{itemize}
\item[A1)] 
 $\mathbf{v}_{k,i}$ is temporally and spatially white noise whose covariance matrix is $R_{v_k,i}$ and which is independent of  $\mathbf{U}_{k',i'}$ for all $k'$ and $i'$, with $k,k' \in \{1,2,\ldots,N\}$ and $i,i' >0$;
\item[A2)]  $\mathbf{U}_{k,i}$ is independent of  $\mathbf{U}_{k,i'}$, with $i,i' > 0$ and $i \neq i'$ \ (temporal independence).
\item[A3)]  $\mathbf{U}_{k,i}$ is independent of  $\mathbf{U}_{k',i}$, with $k, k' \in \{1,2, \ldots, N\}$ and $k \neq k'$ \ (spatial independence),
\item[A4)] $\mathbf{U}_{k_g,i}$,  $\mathbf{U}_{k_{jc},i}$ and $\mathbf{U}_{k_l,i}$ are independent for all $k \in \{1,2, \ldots, N\}$ and $j\in \{1,2, \ldots, J\}$;
\end{itemize}
In order to evaluate the fourth-order moment of the matrix-valued regression data in Subsection~\ref{subsec:subsecSteadyState}, we further assume:
\begin{itemize}
\item[A5)]  $\mathbf{U}_{k,i}$  $(L_k \times M_k)$ has a real matrix variate normal distribution specified by mean $0_{L_k \times M_k}$ and positive-semidefinite matrices $\Psi_k$ $(M_k\times M_k)$ and $\Omega_k$ $(L_k \times L_k)$ (see~\cite[Chapter 2]{gupta1999matrix}). Equivalently, using standard notation for multivariate normal distribution, the distribution of $\mathbf{U}_{k,i}$ can be defined as $\mathrm{vec}(\mathbf{U}_{k,i}) \sim \mathcal{N}_{_{M_k L_k}} \big(\mathrm{vec}(0_{L_k \times M_k}), \Psi_k \otimes \Omega_k \big)$.
\end{itemize}
\emph{Remark 1:} Note that even for the vector-valued regression data, in order to evaluate the fourth-order moment, the Guassian assumption is required (e.g. see~\cite{lopes2008diffusion} and~\cite{cattivelli2010diffusion}). The results of the fourth-order moment of the matrix-valued regression data appear to be quite a bit more challenging than those on its vector counterpart, due to the extra dimension involved. Therefore, the assumption A5) seems well-justified. 

\subsection{Mean stability }\label{subsec:subsecMeanStability}

By taking the expectation of~\eqref{eps:eq_diff_perf_12} and using assumptions (A1-A3), we obtain
\begin{gather}\label{eps:eq_mean_1}
\begin{split}
E \boldsymbol{\tilde{q}}_{i}= \breve{\mathcal{A}} \, (I-  \mathcal{M}\mathcal{R}_U)   \, \breve{\mathcal{C}} \, E \boldsymbol{\tilde{q}}_{i-1},
\end{split}
\end{gather}
where 
\begin{gather}\label{eps:eq_mean_2}
\begin{split}
\mathcal{R}_U  = E\boldsymbol{\mathcal{D}}_i = \mathrm{blockdiag} \{ R_{U_1} ,  R_{U_2} , \ldots, R_{U_N} \},
\end{split}
\end{gather}
and  
\begin{gather}\label{eps:eq_mean_2_2}
\begin{split}
R_{U_k} = E \, \mathbf{U}_{k,i}^H\mathbf{U}_{k,i}= \mathrm{blockdiag} \{ R_{U_{k_{g}}} ,  R_{U_{k_{jc}}} , R_{U_{k_{l}}} \}.
\end{split}
\end{gather}

The algorithm in~\eqref{eps:eq_diff_perf_12} is asymptotically unbiased, i.e, $E \boldsymbol{\tilde{q}}_{i} \rightarrow 0_{\breve{M}\times 1}$ as $i  \rightarrow \infty$, if the matrix $\breve{\mathcal{A}}  (I-  \mathcal{M}\mathcal{R}_U) \breve{\mathcal{C}}$ is stable. In order to prove its stability, we will build on the approaches taken in~\cite{takahashi2010diffusion} and~\cite{sayed2012diffusion}, by selecting a convenient matrix norm $\|\cdot\|$ and exploit its submultiplicativity property, i.e., $\|A B\| \leq \|A\| \|B\| $.

Here, we use the induced block maximum matrix norm~\cite{takahashi2010diffusion},~\cite{sayed2012diffusion}, however, defined over a block matrix with different block sizes. In particular, let $x$ be a   $\breve{M} \times 1$ vector consisting of $\breve{N}$ blocks, where $\breve{N}= N + \sum_{k=1}^{n}|\mathcal{I}_k| +N$,  given as
\begin{gather*}\label{eps:eq_diff_perf_65646}
\begin{split}
x= \mathrm{col}\left \{ \left \{x_{k}^{(w)}, \{x_{k}^{(\varsigma_j)}\}_{j \in \mathcal{I}_k },x_{k}^{(\xi)}, \right \}_{k=1}^N \right \}.
\end{split}
\end{gather*}
The block maximum norm is defined by
\begin{gather*}\label{eps:eq_mean_3}
\begin{split}
\|x\|_{b,\infty} = \underset{1 \leq k \leq N}{\mathrm{max}} \left \{ \{ \| x_k^{(w)}\|, \left \{ \| x_k^{(\varsigma_j)}\| \right \}_{j \in \mathcal{I}_k}, \| x_{k}^{(\xi_k)} \|  \right \}
\end{split}
\end{gather*}
where $\|\cdot\|$ denotes the Euclidean norm of its argument. Next, we define the matrix norm induced from the block
maximum norm, i.e.,
\begin{gather*}\label{eps:eq_mean_3}
\begin{split}
\|A\|_{b,\infty} = \, \underset{\|x\|_{b,\infty} =1}{\mathrm{max}} \, \|A x\|_{b,\infty} 
\end{split}
\end{gather*}
where $A$ is $\breve{M} \times \breve{M}$ matrix. As in~\cite{takahashi2010diffusion}, it can be straightforwardly shown that the block maximum
norm has the unitary invariance property of the Euclidean norm under properly defined block-wise transformation.

Next, by evaluating the block maximum norm of~\eqref{eps:eq_mean_1} and  by applying its submultiplicativity property, we obtain the following relation
\begin{gather}\label{eps:eq_mean_4}
\begin{split}
\| E \boldsymbol{\tilde{q}}_{i}\|_{b,\infty} &\leq \|\breve{\mathcal{A}}\|_{b,\infty} \, \|I-  \mathcal{M}\mathcal{R}_U\|_{b,\infty}   \, \|\breve{\mathcal{C}}\|_{b,\infty} \, \|E \boldsymbol{\tilde{q}}_{i-1}\|_{b,\infty}. 
\end{split}
\end{gather}
Let us now evaluate the block maximum norms of the extended combination matrices $\breve{\mathcal{A}}$ and $\breve{\mathcal{C}}$  given in~\eqref{eps:eq_EWM_P1}, e.g., $\|\breve{\mathcal{A}}\|_{b,\infty}$, while the same holds for $\|\breve{\mathcal{C}}\|_{\infty}$. Since $\breve{\mathcal{A}}$ is a right stochastic matrix, we can bound  $\|\breve{\mathcal{A}}\|_{\infty}$ as shown in~\eqref{eps:eq_Appendix_norms_6} at the top of the next page.
\begin{figure*}[!ht]
\begin{gather}\label{eps:eq_Appendix_norms_6}
\begin{split}
\|  \breve{\mathcal{A}}x\|_{b,\infty} &= \underset{1\leq k \leq N}{ \mathrm{max} } \left \{ \left\| \sum_{\ell \in \mathcal{N}_k } a_{k,\ell}^w x_\ell^{(w)} \right\|, \left \{ \left\| \sum_{\ell \in \mathcal{N}_k \cap \mathcal{C}_j} a_{k,\ell}^{\varsigma_j} x_\ell^{(\varsigma_j)} \right\| \right\}_{j \in \mathcal{I}_k}, \left\| x_{k}^{(\xi_k)} \right\|  \right \} \\
& \leq \underset{1\leq k \leq N}{ \mathrm{max} } \left \{ \sum_{\ell \in \mathcal{N}_k } | a_{k,\ell}^w |, \underset{j \in \mathcal{I}_k}{ \mathrm{max} } \sum_{\ell \in \mathcal{N}_k \cap \mathcal{C}_j} | a_{k,\ell}^{\varsigma_j} | , 1  \right \} \|x\|_{b,\infty} = \|x\|_{b,\infty}.
\end{split}
\end{gather}
\rule{\textwidth}{0.1mm}
\end{figure*} 
Thus, $\|\breve{\mathcal{A}}\|_{b,\infty} < 1$, given that $A^{w}$, $A^{\varsigma_j}$ are row-stochastic, i.e., $A^{w}1_{N}=1_{N}$ and  $A^{\varsigma_j}1_{|\mathcal{C}_j|}=1_{|\mathcal{C}_j|} $, for $j=\{1, \ldots , J\}$.

At this point, we only need to find the conditions that secure 
\begin{gather*}\label{eps:eq_mean_5}
\begin{split}
 \|I-  \mathcal{M}\mathcal{R}_U\|_{b,\infty} < 1.
\end{split}
\end{gather*}
Under assumption A4, due to the unitary invariance of  the block maximum
these conditions correspond to the mean stability conditions of stand-alone LMS filters and can be easily realized to be 
\begin{gather*}\label{eps:eq_mean_6}
\begin{split}
 \mu_k < \frac{2}{\lambda_{\mathrm{max}}(R_{U_{k_{g}}} ,  R_{U_{k_{jc}}} , R_{U_{k_{l}}})}  \qquad \text{for each} \, \, k, 
\end{split}
\end{gather*}
where $k=\{1, \ldots , N\}$ and $\lambda_{\mathrm{max}}(X,Y,Z)$ denotes the maximum of the maximum eigenvalues of the Hermitian matrix arguments $X,Y$ and $Z$.

The above discussion is summarized in the subsequent theorem.
\begin{theorem}\label{teo:teo1} 
For any initial conditions, under the assumptions A1-A4 made in Subsection~\ref{subsec:subsecDataAssumptions}, if the positive step-size of each node satisfies $ \mu_k < 2/\lambda_{\mathrm{max}}(R_{U_{k_{g}}} ,  R_{U_{k_{jc}}} , R_{U_{k_{l}}})$, then
the estimates generated by ATC (or CTA) D-NSPE algorithm converge in the mean, i.e., \\
\begin{gather}\label{eps:eq_teo1}
\begin{split}
\underset{i\rightarrow \infty}{\mathrm{lim}} E \boldsymbol{\tilde{q}}_{i}=0_{\breve{M} \times 1},
\end{split}
\end{gather}
if the combination matrices related to the estimates of global and common parameters are row-stochastic.
\end{theorem}

\subsection{Steady-state performance}\label{subsec:subsecSteadyState}

At this point, we aim to evaluate the mean-square performance of the general diffusion model in~\eqref{eps:eq_diff_perf_12}. In particular, we will examine the performance in the stady-state in terms of MSD and EMSE.

To this end, we use the energy conservation arguments~\cite{sayed2011adaptive},~\cite{sayed2012diffusion}. Specifically, after equating the weighted norm of~\eqref{eps:eq_diff_perf_12} and taking the expectation under Assumptions A1-A3, we obtain the subsequent variance relation
\begin{gather}\label{eps:SteadyState_eq_1}
\begin{split}
E \|\boldsymbol{\tilde{q}}_{i} \|_{\Sigma}^2=E \|\boldsymbol{\tilde{q}}_{i-1} \|_{\Sigma'}^2 +   E \left \{ \boldsymbol{\mathcal{V}}_i^H  \,\mathcal{M}\breve{\mathcal{A}}^T \,\Sigma \, \breve{\mathcal{A}} \,\mathcal{M} \boldsymbol{\mathcal{V}}_i \right\}
\end{split}
\end{gather}
where $\Sigma$ is an arbitrary $(\breve{M} \times \breve{M})$ Hermitian nonnegative-definite matrix that we are free to choose, and 
\begin{gather}\label{eps:SteadyState_eq_2}
\begin{split}
\Sigma'= E \left \{ \breve{\mathcal{C}}^T \, (I-  \mathcal{M}\boldsymbol{\mathcal{D}}_i)^H   \,\breve{\mathcal{A}}^T \, \Sigma   \, \breve{\mathcal{A}} \, (I-  \mathcal{M}\boldsymbol{\mathcal{D}}_i)   \,\breve{\mathcal{C}} \right \}.
\end{split}
\end{gather}

To proceed, we have to extract $\Sigma$ from r.h.s. of~\eqref{eps:SteadyState_eq_2} and from the second term on r.h.s. in~\eqref{eps:SteadyState_eq_1}. To do so, we will use vectorization operator and exploit some useful properties of the trace operator and Kronecker product, i.e.,
\begin{gather}\label{eps:Kron_eq1}
\begin{split}
\mathrm{vec}(ABC)= (C^T \otimes A) \mathrm{vec}(B)
\end{split}
\end{gather}
and
\begin{gather}\label{eps:Kron_eq2}
\begin{split}
\mathrm{Tr} (AB) = \mathrm{vec}(A^T)^T \mathrm{vec}(B).
\end{split}
\end{gather}
Furthermore, in addition to Assumptions A1-A3, here we also use Assumption A5, stated in Subsection~\ref{subsec:subsecDataAssumptions}.

Thus, after defining $ V= E \,\boldsymbol{\mathcal{V}}_i \boldsymbol{\mathcal{V}}_i^H=\mathrm{blockdiag}\left( \left \{ (\mathrm{Tr}  \,R_{v,k} \Omega_k) \Psi_k \right \}_{k=1}^{N} \right)$~\cite{neudecker1987fourth}, we get
\begin{gather}\label{eps:SteadyState_eq_3}
\begin{split}
E \|\boldsymbol{\tilde{q}}_{i} \|_{\Sigma}^2=E \|\boldsymbol{\tilde{q}}_{i-1} \|_{\Sigma'}^2 + \mathrm{Tr} \left(  \breve{\mathcal{A}} \,\mathcal{M} \,V \,\mathcal{M}\breve{\mathcal{A}}^T \,\Sigma \right) .
\end{split}
\end{gather}
Next, we introduce $\sigma=\mathrm{vec}(\Sigma)$. In order to extract $\Sigma $ from $\Sigma'$, we take the following steps
\begin{gather}\label{eps:SteadyState_eq_4}
\begin{split}
\sigma'=\mathrm{vec}(\Sigma')= F \sigma
\end{split}
\end{gather}
where $F$ is a matrix, of dimensions $\breve{M}^2 \times \breve{M}^2$, given by
\begin{gather}\label{eps:SteadyState_eq_5}
\begin{split}
F &= E \left \{ \left( \breve{\mathcal{A}} \, (I-  \mathcal{M}\boldsymbol{\mathcal{D}}_i)   \,\breve{\mathcal{C}}\right)^T \otimes \left(\breve{\mathcal{C}}^T \, (I-  \mathcal{M}\boldsymbol{\mathcal{D}}_i)^T   \,\breve{\mathcal{A}}^T \right) \right \} \\
&= (\breve{\mathcal{C}}^T  \otimes \breve{\mathcal{C}}^T)  E \left \{ (I-  \mathcal{M}\boldsymbol{\mathcal{D}}_i)^T \breve{\mathcal{A}}^T \otimes ((I-  \mathcal{M}\boldsymbol{\mathcal{D}}_i)^T   \breve{\mathcal{A}}^T  \right \} \\     
&= (\breve{\mathcal{C}}^T  \otimes \breve{\mathcal{C}}^T) \, G \, (\breve{\mathcal{A}}^T  \otimes \breve{\mathcal{A}}^T)
\end{split}
\end{gather}
with
\begin{gather}\label{eps:SteadyState_eq_6}
\begin{split}
G =&  E \left \{ (I-  \mathcal{M}\boldsymbol{\mathcal{D}}_i)^T \otimes (I-  \mathcal{M}\boldsymbol{\mathcal{D}}_i)^T  \right \}\\
=& I\otimes I - \mathcal{R}_U \mathcal{M} \otimes I - I \otimes \mathcal{R}_U \mathcal{M} + E\{ \boldsymbol{\mathcal{D}}_i^T \mathcal{M} \otimes \boldsymbol{\mathcal{D}}_i^T \mathcal{M} \}
\end{split}
\end{gather}
and $\mathcal{R}_U=\mathrm{blockdiag}\left( \left \{ \mathrm{Tr} ( \Omega_k) \Psi_k \right \} \right)$ (see~\eqref{eps:eq_mean_2}).

For sufficiently small step sizes, the forth-order moment of regressors, i.e., the rightmost term in~\eqref{eps:SteadyState_eq_6}, can be discarded. However, this term can be evaluated as follows 
\begin{gather}\label{eps:SteadyState_eq_7}
\begin{split}
 E\{ \boldsymbol{\mathcal{D}}_i^T \mathcal{M} \otimes \boldsymbol{\mathcal{D}}_i^T \mathcal{M} \} &= S \, \cdot (\mathcal{M} \otimes  \mathcal{M})
\end{split}
\end{gather}
where
\begin{gather}\label{eps:SteadyState_eq_7_1}
\begin{split}
S &= E\{ \boldsymbol{\mathcal{D}}_i^T  \otimes \boldsymbol{\mathcal{D}}_i^T  \} \\
&=\mathrm{blockdiag}\left( \left \{ E \{ \mathbf{U}_{k,i}^H \mathbf{U}_{k,i}  \otimes \boldsymbol{\mathcal{D}}_i \} \right \}_{k=1}^{N} \right )
\end{split}
\end{gather}
with
\begin{gather}\label{eps:SteadyState_eq_7_1a}
\begin{split}
 E &\{ \mathbf{U}_{k,i}^H \mathbf{U}_{k,i}  \otimes \boldsymbol{\mathcal{D}}_i \}= K_{(M_k, \breve{M})}  E \{ \boldsymbol{\mathcal{D}}_i \otimes  \mathbf{U}_{k,i}^H \mathbf{U}_{k,i}  \} K_{(\breve{M}, M_k)} 
\end{split}
\end{gather}
and $K_{(m, n)}$ denoting the $mn \times mn$ commutation matrix that satisfies 
\begin{gather*}
\begin{split}
K_{(m,n)} \mathrm{vec}(A) = \mathrm{vec}(A^T) 
\end{split}
\end{gather*} 
for any $m \times n$ matrix $A$~\cite{{magnus1979commutation}}. In~\eqref{eps:SteadyState_eq_7_1a}, it can be shown that
\begin{gather}\label{eps:SteadyState_eq_7_1b}
\begin{split}
E \{ &\boldsymbol{\mathcal{D}}_i \otimes  \mathbf{U}_{k,i}^H \mathbf{U}_{k,i}  \}\\ 
&  = \mathrm{blockdiag} \left (  \left \{ E \{ \mathbf{U}_{\ell,i}^H \mathbf{U}_{\ell,i}  \otimes \mathbf{U}_{k,i}^H \mathbf{U}_{k,i} \} \right \}_{\ell=1}^{N}\right ).
\end{split}
\end{gather}
Moreover, from~\cite{neudecker1987fourth} and~\cite{von1988moments}, we can obtain closed-form expressions for the the expectations that appear in~\eqref{eps:SteadyState_eq_7_1a}. In particular, we can check that
\begin{gather}\label{eps:SteadyState_eq_7_1c}
\begin{split}
 E \{ \mathbf{U}_{k,i}^H \mathbf{U}_{k,i} & \otimes \mathbf{U}_{k,i}^H \mathbf{U}_{k,i} \} \\ &= \mathrm{Tr}( \Omega_k) \mathrm{Tr}( \Omega_k) \Psi_k \otimes \Psi_k\\
& \phantom{=} +\mathrm{Tr}( \Omega_k  \Omega_k ) \mathrm{vec}( \Psi ) \mathrm{vec}( \Psi ) ^T\\
& \phantom{=} +\mathrm{Tr}( \Omega_k  \Omega_k ) K_{(M_k, M_k)} (\Psi_k  \otimes \Psi_k )
\end{split}
\end{gather}
and that
\begin{gather}\label{eps:SteadyState_eq_7_1d}
\begin{split}
 E \{ \mathbf{U}_{\ell,i}^H \mathbf{U}_{\ell,i}  \otimes \mathbf{U}_{k,i}^H \mathbf{U}_{k,i} \} =  \left [ \mathrm{Tr} ( \Omega_\ell) \Psi_\ell \right ] \otimes \left [  \mathrm{Tr} ( \Omega_k) \Psi_k  \right ]
\end{split}
\end{gather}
for any $k, \ell \in \{1,2,\ldots,N\}$ with $k \neq \ell$.

To evaluate the performance measures in the steady state, i.e., $i \to \infty$, by using~\eqref{eps:Kron_eq2}, we first rewrite~\eqref{eps:SteadyState_eq_3} as
\begin{gather}\label{eps:SteadyState_eq_8}
\begin{split}
E \|\boldsymbol{\tilde{q}}_{\infty} \|_{\sigma}^2=E \|\boldsymbol{\tilde{q}}_{\infty} \|_{F\sigma}^2 +  \left[ \mathrm{vec}(\breve{\mathcal{A}} \,\mathcal{M} \,V^T \,\mathcal{M}\breve{\mathcal{A}}^T \,) \right]^T \sigma.
\end{split}
\end{gather}
After rearranging, we obtain the following relation
\begin{gather}\label{eps:SteadyState_eq_9}
\begin{split}
E \|\boldsymbol{\tilde{q}}_{\infty} \|_{(I-F)\sigma}^2= \left[ \mathrm{vec}(\breve{\mathcal{A}} \,\mathcal{M} \,V^T \,\mathcal{M}\breve{\mathcal{A}}^T \,) \right]^T \cdot \sigma.
\end{split}
\end{gather}
Now, we can evaluate MSD averaged across the whole network
\begin{gather}\label{eps:SteadyState_eq_10}
\begin{split}
\mathrm{MSD}^{\mathrm{net}}= \frac{1}{N} E \|\boldsymbol{\tilde{q}}_{\infty} \|_{I}^2 
\end{split}
\end{gather}
by selecting $\sigma=(I-F)^{-1} \frac{1}{N}\mathrm{vec}(I_{\breve{M}})$, we obtain
\begin{gather}\label{eps:SteadyState_eq_11}
\begin{split}
\mathrm{MSD}^{\mathrm{net}}= \frac{1}{N} & \left[ \mathrm{vec}(\breve{\mathcal{A}} \,\mathcal{M} \,V^T \,\mathcal{M}\breve{\mathcal{A}}^T \,) \right]^T (I-F)^{-1}  \mathrm{vec}(I_{\breve{M}}).\\
\end{split}
\end{gather}
Now, in order to evaluate MSD at each node $k$, let us first define the \emph{Khatri-Rao} matrix product.

\emph{Definition 2:} Consider matrices $A$ and $B$ of dimensions $m \times n$ and  $p \times q$, respectively. Let $A=(A_{ij})$ be partitioned with $A_{ij}$ of dimensions $m_i \times n_j$ as the $(i,j)$-th block submatrix and let $B=(B_{ij})$ be partitioned with $B_{ij}$ as the $(i,j)$-th block submatrix of dimensions $p_i \times q_j$  ($\sum m_i=m$, $\sum n_j=n$, $\sum p_i=p$ and $\sum q_j=q$). The Khatri-Rao matrix product is defined as
\begin{gather*}
\begin{split}
A \odot B= (A_{ij} \otimes B_{ij})_{ij}
\end{split}
\end{gather*}
where  $A_{ij} \otimes B_{ij}$ is of dimensions $m_i p_i \times n_j q_j$,  while $A \odot B$ is of dimensions  $(\sum m_i p_i) \otimes \sum n_j q_j)$, (see~\cite{liu1999matrix}).

Based on the previous definition, MSD at node $k$ is
\begin{gather}\label{eps:SteadyState_eq_12}
\begin{split}
\mathrm{MSD}_{k}=  \left[ \mathrm{vec}(\breve{\mathcal{A}} \,\mathcal{M} \,V^T \,\mathcal{M}\breve{\mathcal{A}}^T \,) \right]^T (I-F)^{-1} m_k ,
\end{split}
\end{gather}
where
\begin{gather}\label{eps:SteadyState_eq_13}
\begin{split}
m_{k}= \mathrm{vec}\left(\mathrm{diag}(e_k) \odot Y \right)
\end{split}
\end{gather}
with $N \times N$ partitioned matrices $Y=\mathrm{blockdiag}(I_{M_1},\ldots, I_{M_N})$ and block-diagonal matrix made of the elements of a $1 \times N$ vector $e_k$ with the unity at the $k^{th}$ position and zeros elsewhere.

On the other hand, MSD related to the estimation of the global, some specific common or the local vector of parameter at node $k$ can be evaluated by redefining $Y$ as a $(2N+\sum^N_{k=1}|\mathcal{I}_k|) \times (2N+\sum^N_{k=1}|\mathcal{I}_k|)$ partitioned matrix, i.e.,
\begin{gather}\label{eps:SteadyState_eq_14}
\begin{split}
Y=\mathrm{blockdiag}(I_{M_g}, I_{|\mathcal{I}_1|M_c},I_{M_l},\ldots, I_{M_g},I_{|\mathcal{I}_N|M_c},I_{M_l}),
\end{split}
\end{gather}
and by taking $1 \times 2N+\sum^N_{k=1}|\mathcal{I}_k|$ vector $e_k$ with the unity at the appropriate position and zeros elsewhere.

Similarly,
\begin{gather}\label{eps:SteadyState_eq_15}
\begin{split}
\mathrm{EMSE}^{\mathrm{net}}= \frac{\left[ \mathrm{vec}(\breve{\mathcal{A}} \,\mathcal{M} \,V^T \,\mathcal{M}\breve{\mathcal{A}}^T \,) \right]^T (I-F)^{-1} \mathrm{vec}(\mathcal{R}_U)}{N}.
\end{split}
\end{gather}
Additionally, EMSE at each node $k$ is
\begin{gather}\label{eps:SteadyState_eq_12}
\begin{split}
\mathrm{EMSE}_{k}= & \left[ \mathrm{vec}(\breve{\mathcal{A}} \,\mathcal{M} \,V^T \,\mathcal{M}\breve{\mathcal{A}}^T \,) \right]^T (I-F)^{-1} p_k ,
\end{split}
\end{gather}
where we select a node $k$ by 
\begin{gather}\label{eps:SteadyState_eq_13}
\begin{split}
p_{k}= \mathrm{vec}\left(\mathrm{diag}(e_k) \odot \mathcal{R}_U \right)
\end{split}
\end{gather}
where $\mathcal{R}_U$ is defined as $N \times N$ partitioned matrix as in~\eqref{eps:eq_mean_2}. Under the independence of $ \mathbf{U}_{k_g,i}$, $\mathbf{U}_{k_{jc},i}$ and $  \mathbf{U}_{k_l,i}$, we can evaluate EMSE performance measure related to the global, specific common or local parameter at some node $k$. To do so, we need to properly redefine the partitions of $R_U$ and the size of vector $e_k$.

\section{Simulation results}
\label{sec:sec6}

In this section, we initially discuss some generic simulations that verify mean-square theoretical results (see Section~\ref{subsec:subsecSteadyState}). Afterwards, the effectiveness of the proposed algorithms are illustrated in the context of cooperative spectrum sensing in CR networks. 

\subsection{Validation of mean-square theoretical results}

\begin{figure}[t]
  \centering
\centerline{\epsfig{figure=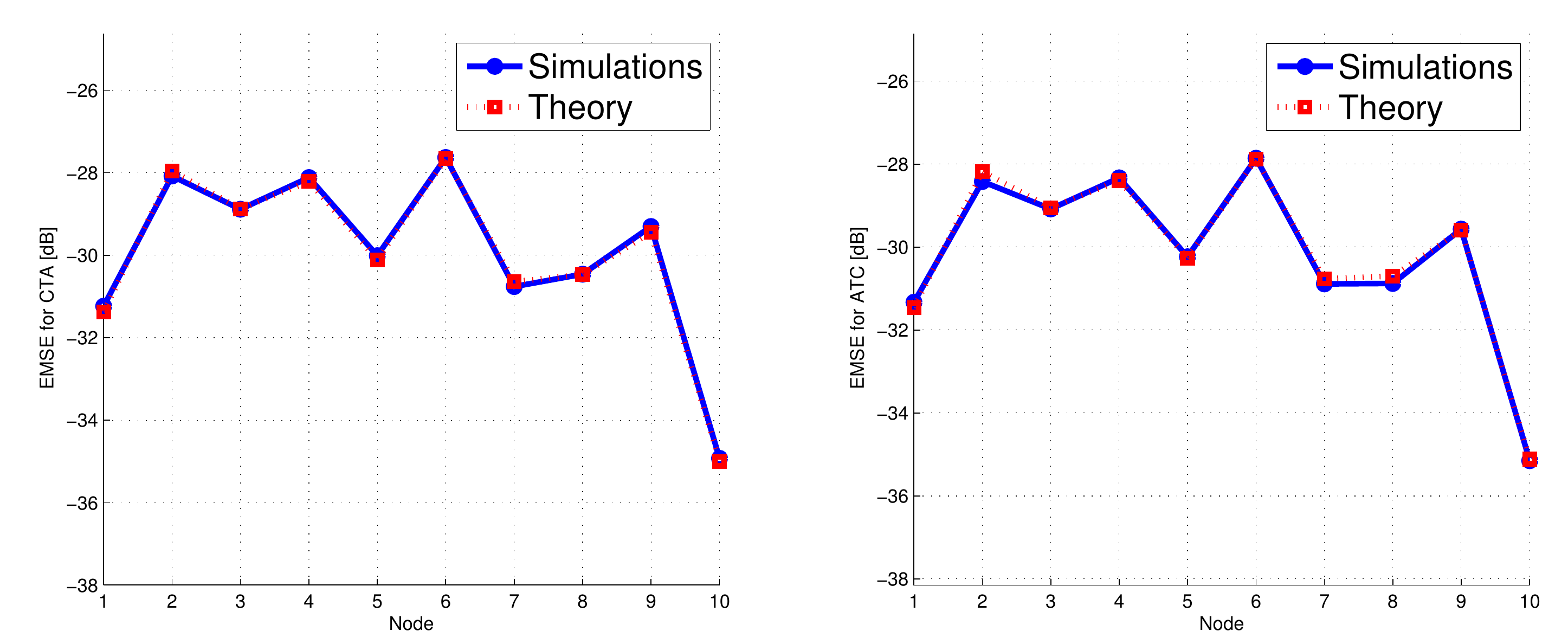,width=0.475\textwidth}}
\caption{Steady-state EMSE per node for CTA D-NSPE and ATC D-NSPE.}
\label{fig:ResultsEMSE}
\end{figure}

We assume a network with $N=10$ nodes where the measurements follow the observation model provided in~\eqref{eps:eq3} with $L_k=2$ for all $k$. In the considered setting, two different vectors of common parameters coexist, i.e., $\varsigma_1^{o}$ and $\varsigma_2^{o}$. The vector $\varsigma_1^{o}$ is composed of 3 parameters, while $\varsigma_2^{o}$ consists of 2 parameters. Moreover, we consider that the area of influence of $\varsigma_1^{o}$ and $\varsigma_2^{o}$ is formed by $\mathcal{C}_1 \in \{2,3,4,5,6\}$ and $\mathcal{C}_2 \in \{5,6,7,8\}$, respectively. As a result, there are nodes that are interested in estimating zero, one or two different vector of common parameters. In addition, each node $k$ is interested in estimating a vector of global parameters and a vector of local parameters, each one of length equal to $M_g=2$ and $M_{l_k}=3$, respectively. 

The data observed by each node, i.e., $\{d_{k,i}, U_{k,i}\}$, have been generated under the assumption of a background noise $\mathbf{v}_{k,i}$ with covariance $\sigma_{v_k}^2 I_2$, where $\sigma_{v_k}^2=\sigma_{v}^2=10^{-3}$ across the network. Furthermore, each one of the $L_k$ rows of the regressor 
\begin{displaymath}
U_{k,i} = \mathrm{col} \left \{ U_{k_g,i}^{T} \{U_{k_{jc},i}^{T}\}_{j \in \mathcal{I}_k} U_{k_l,i}^{T} \right \}^T 
\end{displaymath}
have been independently drawn from a time-correlated spatially independent Gaussian distribution. In particular, the $c$-th row of $U_{k,i}$ is generated according to a first-order autoregressive (AR) model with correlation function $r_{k,c}(i)=\sigma_{u_k}^2 \alpha_{k}^{|i|}$ where the pair of parameters $\{\sigma_{u_k}, \alpha_{k}\}$ are randomly chosen in (0,1) so that the the Signal-to-Noise-Ratio (SNR) at each node ranges from 10 dB to 20 dB. Hence, $\mathbf{U}_{k,i}$ follows a real matrix variate normal distribution specified by the mean matrix $0_{2 \times M_k}$ and the positive-semidefinite matrices  $\Omega_k=I_{2}$ and $\Psi_k=\mathrm{toeplitz}\{[\sigma_{u_k}^2 \, \sigma_{u_k}^2 \alpha_k \, \ldots \, \sigma_{u_k}^2 \alpha_{k}^{M_k-1}]^{T}\}$.

When implementing both CTA D-NSPE and ATC D-NSPE, static uniform combination weights have been assumed, i.e., $a_{k,\ell}^{w} = 1/|\mathcal{N}_{k}|$ for all $k,\ell \in \{1,2,\ldots,10\}$, and $a_{k,\ell}^{\varsigma_j} = 1/|\mathcal{N}_k \cap \mathcal{C}_j|$ for all $k,\ell \in \mathcal{C}_j$ and $j \in \{1,2\}$. The neighborhood of each node $k$ has been set so that the network graph as well as the subsets $\mathcal{C}_1$ and $\mathcal{C}_2$ are connected. Moreover, in order to validate the theoretical expressions for non-fully connected networks and non-fully connected subsets $\mathcal{C}_j$, we have assumed that $\underset{1 \leq k \leq N}{\mathrm{max}}\{|\mathcal{N}_k|\} \leq 4$ and that $\underset{1 \leq k \leq N}{\mathrm{max}}\{|\mathcal{N}_k \cap \mathcal{C}_j |\} \leq 2$. 

\begin{figure}[t]
  \centering
\centerline{\epsfig{figure=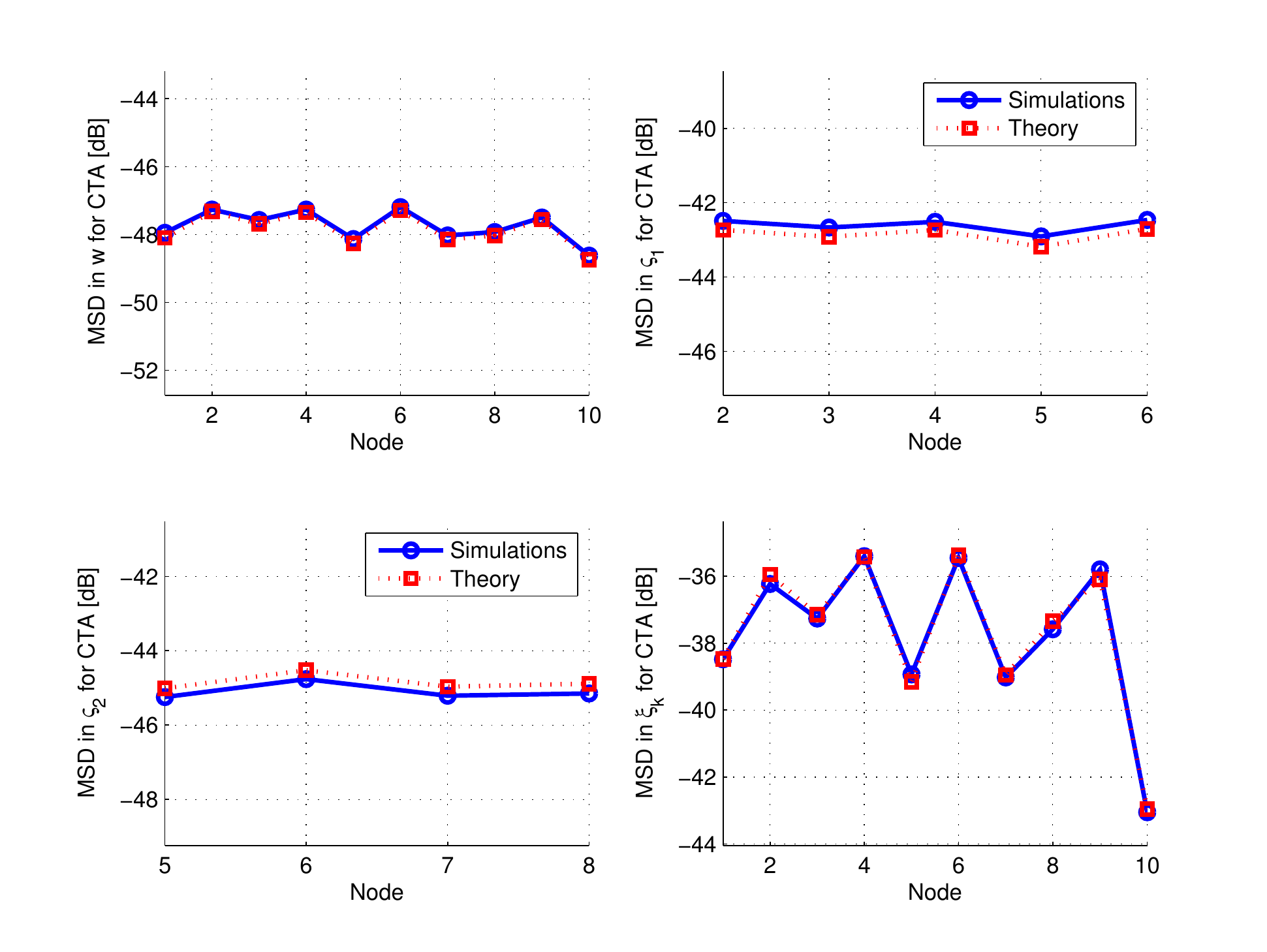,width=7.9cm}}
\caption{Steady-state MSD per node for CTA D-NSPE LMS.}
\label{fig:figMSD_GCCL_CTA}
\end{figure}

\begin{figure}[t]
  \centering
\centerline{\epsfig{figure=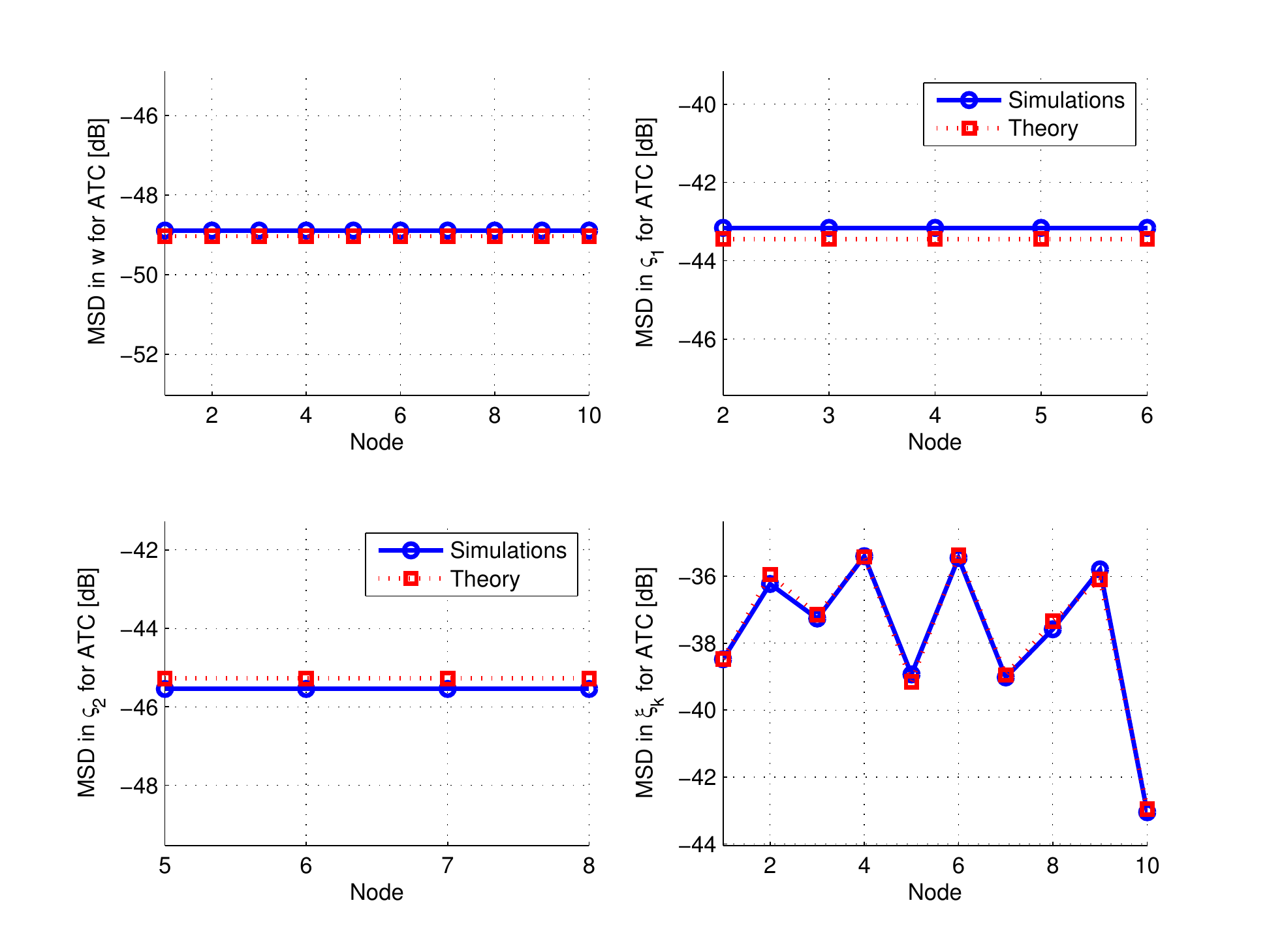,width=7.9cm}}
\caption{Steady-state MSD per node for ATC D-NSPE LMS.}
\label{fig:figMSD_GCCL_ATC}
\end{figure}

The experimental values in~Figs. \ref{fig:ResultsEMSE}-\ref{fig:figMSD_GCCL_ATC} result from averaging the mean-square measures over 100 independent experiments where both CTA D-NSPE LMS and ATC D-NSPE LMS are run for 10 000 iterations. Despite the temporal correlation of the regressors as well as the correlation among $ \mathbf{U}_{k_g,i}$, $\mathbf{U}_{k_{jc},i}$ and $  \mathbf{U}_{k_l,i}$, which was not assumed for the derivation of the theoretical results, all figures show a good match between the simulated curves and the theoretical expressions for the MSD and EMSE at each node $k$. 

\subsection{Illustrative application}

\begin{figure}[t]
  \centering
\centerline{\epsfig{figure=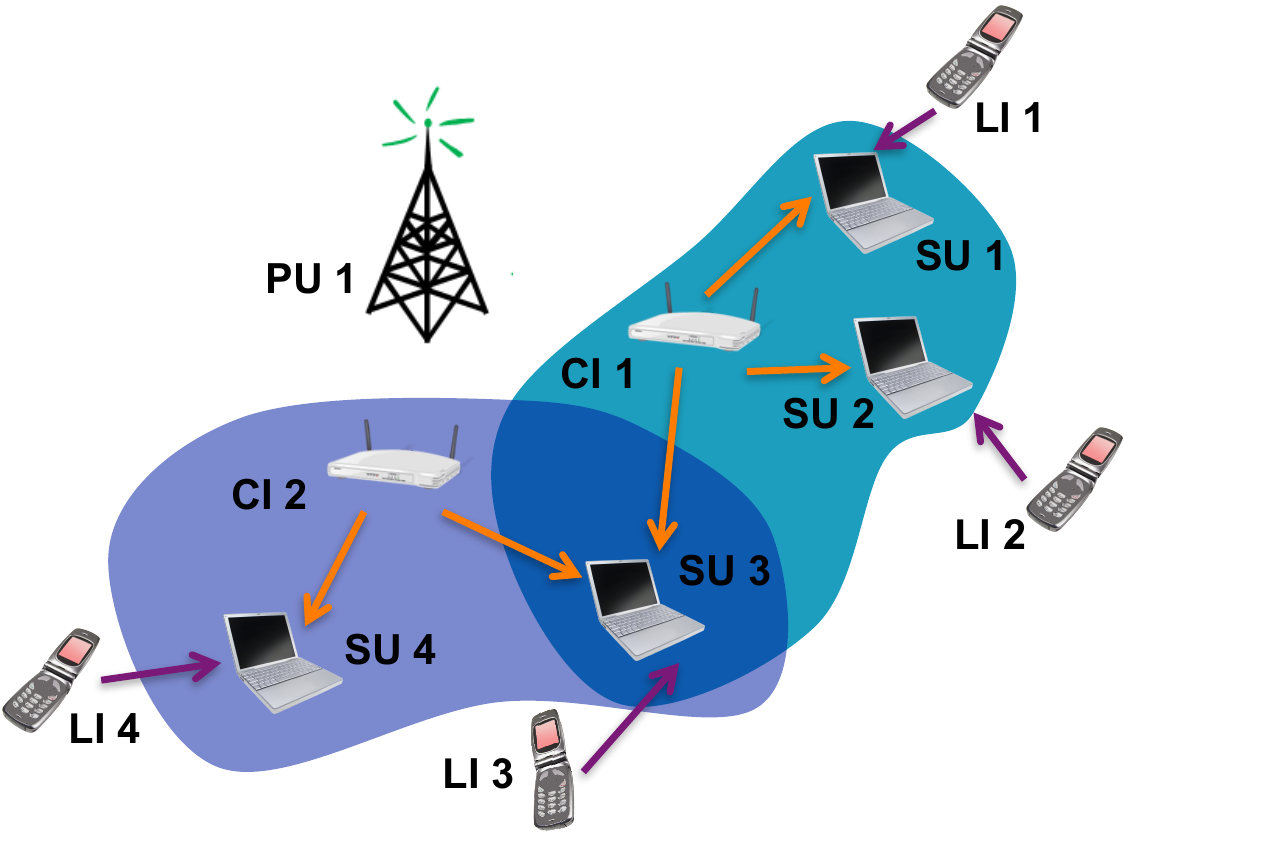,width=7cm}}
\caption{An illustrative CR scenario. Each one of the 4 secondary users (SU) aims at estimating the aggregated spectrum transmitted by the Primary User (PU) and its interferer(s). Apart from the PU and its Local Interferer (LI), each SU is influenced by one or two Common Interferers (CI). SU 3 is influenced by CI 1 and CI 2, while SUs 1, 2  and 4 are only influenced by one CI.}
\label{fig:figCR}
\end{figure}

In the following, we will also demonstrate the performance of the proposed algorithm when used for  cooperative spectrum sensing in CR networks (see~\cite[Section 2.4]{sayed2012diffusion} and~\cite{Di_LorenzoTSP}\nocite{5352337_CR_Giannakis}-\cite{Di_LorenzoLETTERS}). In brief, there are $Q$ primary users (PU) transmitting and $N$ secondary users (SU) sensing the power spectrum. In addition to PUs, for each SU we also assume two types of low-power interference sources, i.e., local interferer (LI) and common interferers (CI). The former is affecting only one SU, while the latter are influencing several SUs. Therefore, the aim for each SU is to estimate the aggregated spectrum transmitted by all the PUs as well as the spectrum of its own LI and CI.
 An example of such a scenario is given in~Fig. \ref{fig:figCR}.

Next, the power spectral density (PSD) of the signal transmitted by the $q$-th PU, denoted by $\Phi_q^t(f)$, can be approximated by using the subsequent model of $X$ basis functions 
\begin{gather}\label{eps:n1}
\begin{split}
\Phi^t_q(f) =\sum_{x=1}^X b_x(f) \check{w}_{qx}^{o} = b_0^T(f) \check{w}_q^{o}
\end{split}
\end{gather}
where $b_0(f)=[ b_1(f), \ldots , b_X(f)]^T \in \mathbb{R}^X$ is a vector of basis functions evaluated at frequency $f$ and $\check{w}_q^{o}=[ \check{w}_{q1}^{o}, \ldots , \check{w}_{qX}^{o}]^{T} \in \mathbb{R}^X$ is a vector of weighting coefficients representing the power transmitted by the $q$-th PU over each basis. 

Let $p_{tk,i}(f)= |H_{tk}(f,i)|^2$ be the frequency-dependent attenuation coefficient, where $H_{tk}(f,i)$ is the channel frequency response between the $t$-th transmitter and $k$-th receiver~\cite{Di_LorenzoLETTERS}. For each time $i$ and frequency $f$, we define
\begin{itemize}
\item[-]$p_{qk,i}(f)$ denoting the attenuation coefficient between the $q$-th PU and the $k$-th SU,
\item[-]$p_{Ik,i}(f)$ refering to the attenuation coefficient between the local interferer and the $k$-th SU,
\item[-]$p_{jk,i}(f)$ being the attenuation coefficient between the $j$-th common interferer and the $k$-th SU, where $j \in \mathcal{I}_k$. 
\end{itemize}
Then, under the assumption of spatial uncorrelation among the channels, the signal received by the $k$-th SU at time instant $i$ can be expressed as 
\begin{gather}\label{eps:n2}
\begin{split}
\Phi^r_{k,i}(f) = b_{k,i}^T(f) w_k^o + z_{k,i},
\end{split}
\end{gather}
where $w_k^{o} = \text{col }\{\check{w}_1^{o}, \ldots , \check{w}_Q^{o}, \varsigma_{\mathcal{I}_k(1)}^{o}, \ldots ,\varsigma_{\mathcal{I}_k(|\mathcal{I}_k|)}^{o}, \xi_k^{o}\} \in \mathbb{R}^{(Q+|\mathcal{I}_k|+1)X} $ with $\xi_k^o$ and  $\varsigma_{j}^o$ equal to the vectors of weighting coefficients representing the power transmitted by the LI and $j$-th CI associated with the $k$-th SU, respectively. Also, $b_{k,i}(f) =p_{k,i}(f) \otimes b_0(f) \in \mathbb{R}^{(Q+|\mathcal{I}_k|+1)X}$,  and 
\begin{gather}\label{eps:n2EXTRA}
\begin{split}
p_{k,i}(f)=[ p_{1k,i}, \ldots , p_{qk,i}, p_{\mathcal{I}_k(1),i}, \ldots , p_{\mathcal{I}_k(|\mathcal{I}_k|),i}, p_{Ik,i}]^{T},
\end{split}
\end{gather}
while $z_{k,i}$ is the measurement and/or model noise. In the above expression, we dropped the frequency index for compactness of notation. Also note that, in practice, the attenuation factors $p_{tk,i}$ cannot be estimated accurately, so we assume access only to noisy estimates $\hat{p}_{tk,i}$ hereafter. 

Considering that, at discrete time $i$, each node $k$ observes the received PSD in~\eqref{eps:n2} over $L$ frequency samples $\{f_m\}_{m=1}^L$, the subsequent vector linear model is obtained
\begin{gather}\label{eps:n3}
\begin{split}
\mathbf{d}_{k,i} &= \mathbf{U}_{k,i} w_k^{o} + \mathbf{v}_{k,i} 
\end{split}
\end{gather}
where $\mathbf{v}_{k,i}$ denotes noise with zero mean and covariance matrix $R_{v_{k}}$ of dimension $L \times L$ and  $\mathbf{U}_{k,i} = \left [ b_{k,i}(f_1) \ldots b_{k,i}(f_L) \right ]^T$ is of dimension $L \times (Q+|\mathcal{I}_k|+1)X$ with $L > (Q+|\mathcal{I}_k|+1)X$. 

\begin{figure}[!t]
\centering 
\subfigure[]{
\includegraphics[width=0.41\textwidth]{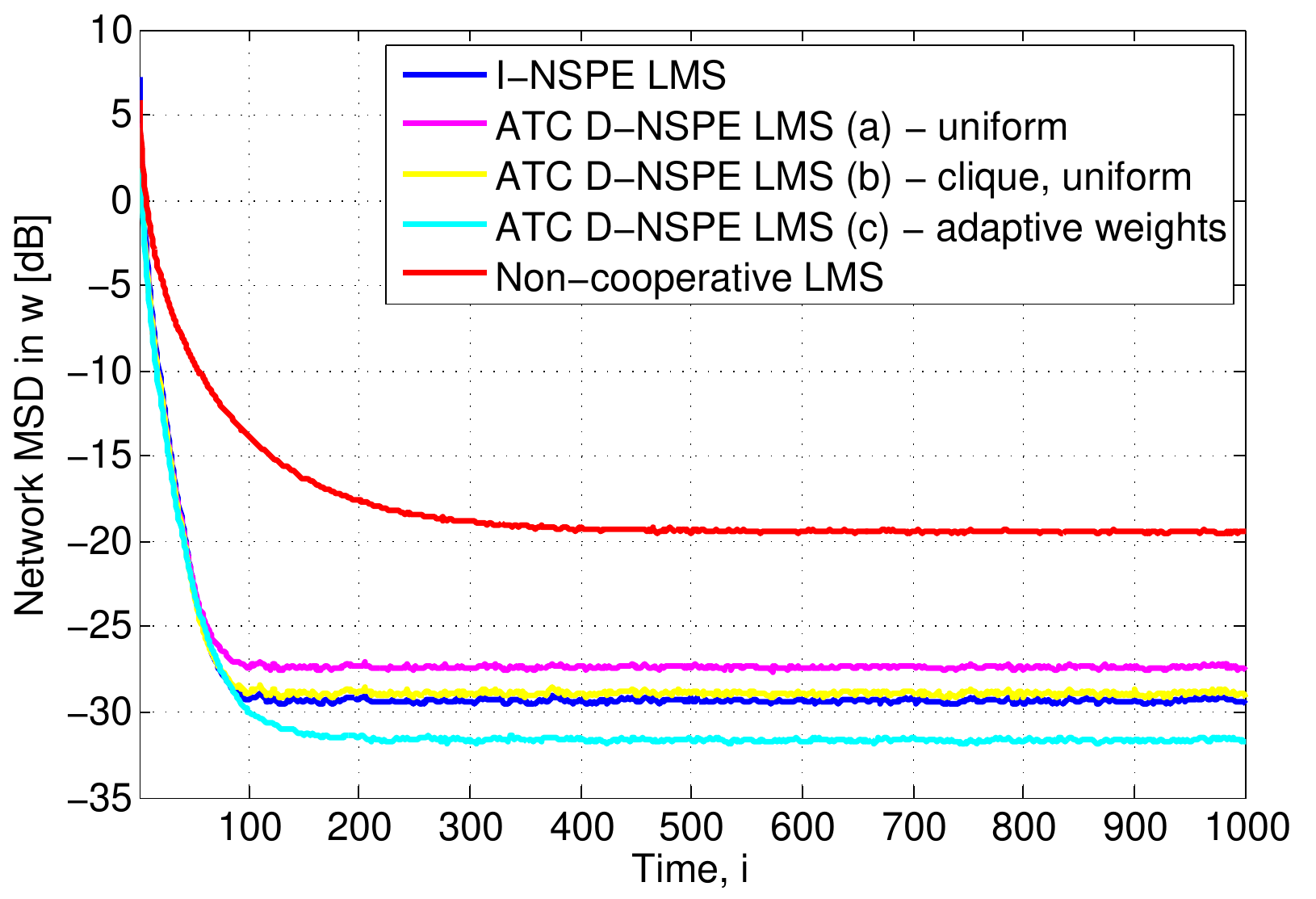} 
\label{fig:fig4a}
} 
\subfigure[]{
\includegraphics[width=0.41\textwidth]{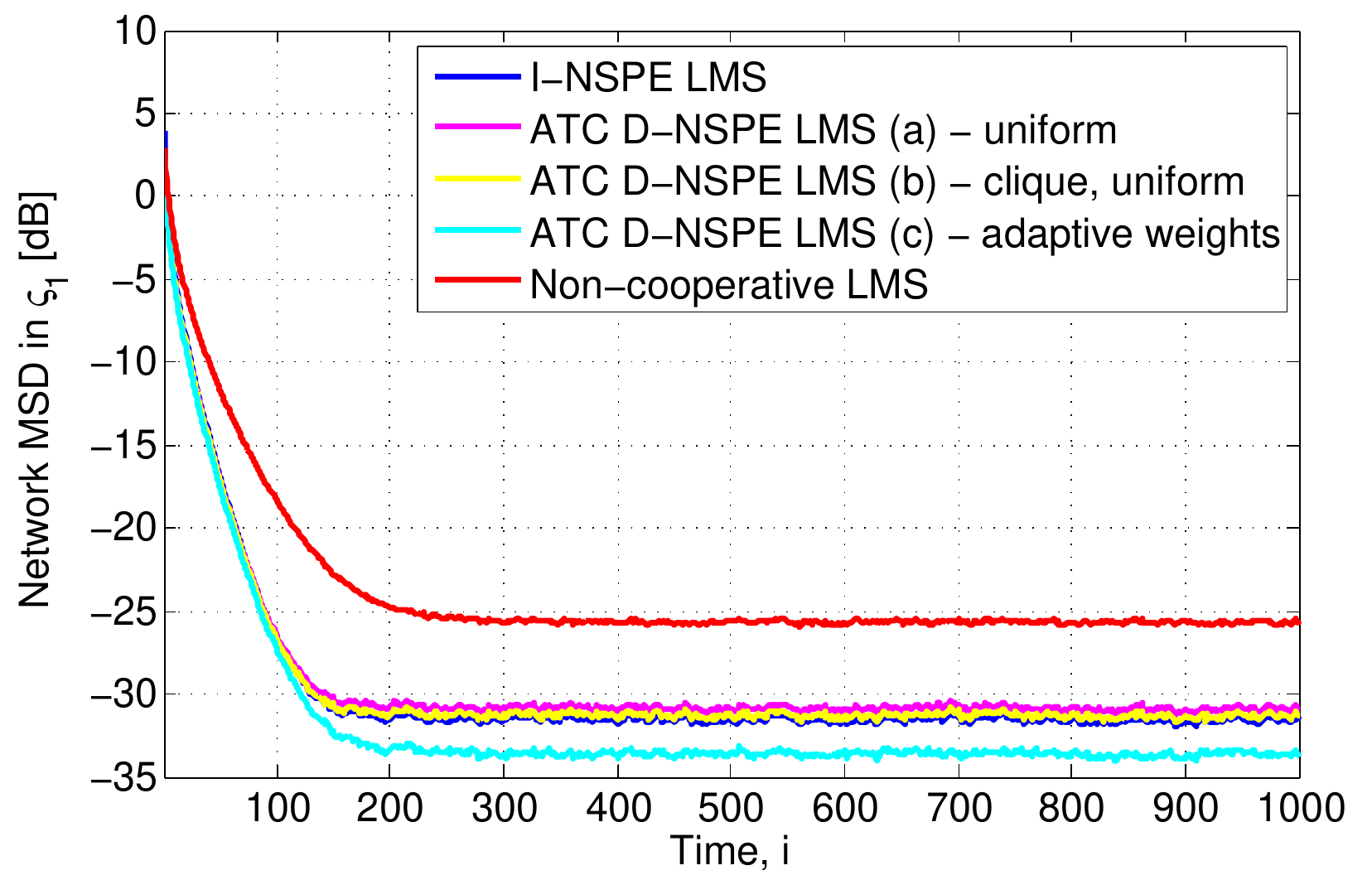} 
\label{fig:fig4b}
}
\subfigure[]{ 
\includegraphics[width=0.41\textwidth]{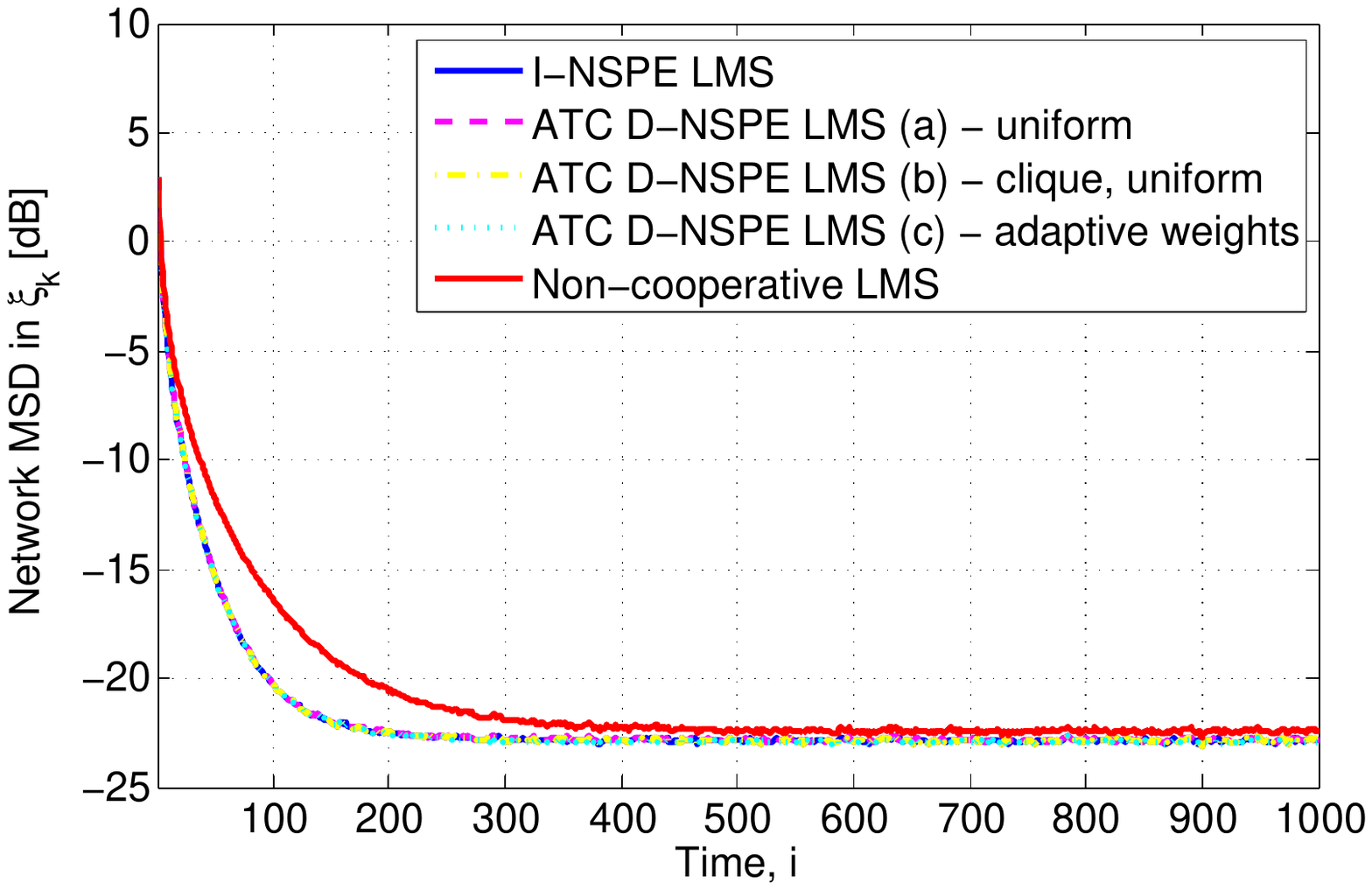}
\label{fig:fig4c}
}
\caption{Learning behavior of network MSD with respect to the parameters of global interest~\subref{fig:fig4a}, common interest~\subref{fig:fig4b} and for the parameters of local interest~\subref{fig:fig4c}.} 
\label{fig:fig4}
\end{figure}

For the computer simulations presented here, we consider a scenario where there is only one common interferer whose PSD can be sensed by nodes in $\mathcal{C}_{1}=\{2, 4, 7, 9\}$. Furthermore, we analyze the ATC D-NSPE LMS scheme for several different combining strategies and degrees of connectivity.
In particular, we consider  the ATC D-NSPE LMS algorithms with
\begin{itemize}
\item[a)] the same neighborhood size at all the nodes, i.e.,  $|\mathcal{N}_{k}|=5$, while  $|\mathcal{N}_{k}\cap\mathcal{C}_1|=3$ for all $k \in {C}_1$. 
In this scenario, we employ the static uniform combination weights, i.e., $a_{k,\ell}^{w}=1/5$ and $a_{k,\ell}^{\varsigma_1}=1/3$.
\item[b)] the clique topology, i.e.,  $|\mathcal{N}_{k}|=N$ and  $|\mathcal{N}_{k}\cap\mathcal{C}_1|=|\mathcal{C}_1|$ for all $k \in {C}_1$, with corresponding static uniform combination weights,
\item[c)] the topology set as in a), while the combination weights are adaptive. Specifically, the weights corresponding to both global and common parameter estimation  processes are being adapted according to the adaptive combination mechanism proposed in~\cite{zhao2012clustering}. For instance, the weights $a_{k,\ell}^{w}$, for $\ell \in \mathcal{N}_{k}$, evolve as
\begin{gather*}
\begin{split}
a_{k,\ell}^{w}(i)=\frac{\gamma_{k,\ell}^{-2}(i)}{\sum_{m \in \mathcal{N}_{k}}^{\phantom{A}}\gamma_{k,m}^{-2}(i)}
\end{split}
\end{gather*}
with
\begin{gather*}
\begin{split}
\gamma_{k,\ell}^{2}(i)=(1-\nu)\gamma_{k,\ell}^{2}(i-1) + \nu ||\psi_{\ell}^{(i)}-\phi_{k,w}^{(i-1)} ||^2.
\end{split}
\end{gather*}
\end{itemize}

We also compare these schemes with an LMS-based non-cooperative strategy as well as with the incremental-based NSPE LMS (I-NSPE LMS), developed in~\cite{bogdanovic2013aj}, that is used as a benchmark. 

The step-size of the LMS adaptation at each node is set equal to $\mu_k=0.04$ for all the algorithms, expect for the incremental NSPE where $\mu_k$ is the step-size for estimating the local parameters only. In the I-NSPE LMS, the step-sizes for estimating global and common parameters are set to $\mu_{w}^{I-NSPE}=\mu_k/N$ and $\mu_{\varsigma_{j}}^{I-NSPE} =\mu_k/|\mathcal{C}_{j}|$, respectively. Accordingly, we obtain a fair comparison among the strategies.

Figure~\ref{fig:fig4} depicts the learning behavior of the two schemes in terms of the network MSD associated with the estimation of $w^o$, $\varsigma_{1}^o$ and $\xi_k^o$. Each network MSD is the result of averaging the local MSDs associated with the estimation of $w^o$ and $\xi_k^o$ at each node, except for the network MSD associated with the estimation of $\varsigma_{1}^o$, which is averaged over the nodes belonging to the set $\mathcal{C}_{1}$. To generate each plot, we have averaged the results over 100 independent experiments where we assumed $Q=2$ PUs, $N=10$ SUs and $X=16$ Gaussian basis functions, of amplitude normalized to one and standard deviation $\sigma_b=0.05$. Furthermore, we have considered that each SU scans $L=80$ channels over the normalized frequency axis between 0 and 1, whereas the noise $z_{k,i}$ in \eqref{eps:n2} is zero-mean Gaussian with standard deviation varying between $0.04$ and $0.16$ for different $k$.

Each attenuation coefficient follows $\hat{p}_{tk,i} (f) = p_{tk,i} (f) + n_{tk}$, where $n_{tk}$ denotes a zero-mean Gaussian variable with standard deviation in the range between $0.3$ and $1.25$, while $p_{tk,i} (f)$ is related to the frequency response of the channel modeled as a static $3$-tap FIR filter. Each tap is assumed to be a zero-mean complex Gaussian random variable with variance $\sigma_h^2=0.25$.
Under this setting, we observe that all the proposed D-NSPE schemes outperform the non-cooperative one, especially when estimating $w^o$ and $\varsigma_{1}^o$. Note that D-NSPE a) and b) well-approximate the centralized-like performance of the incremental strategy. Finally, due to the fact that the adaptive combiners integrate some additional knowledge regarding the quality of the estimates at the different nodes, D-NSPE c) outperforms all other schemes including the incremental.

\begin{figure}[t]
  \centering
\centerline{\epsfig{figure=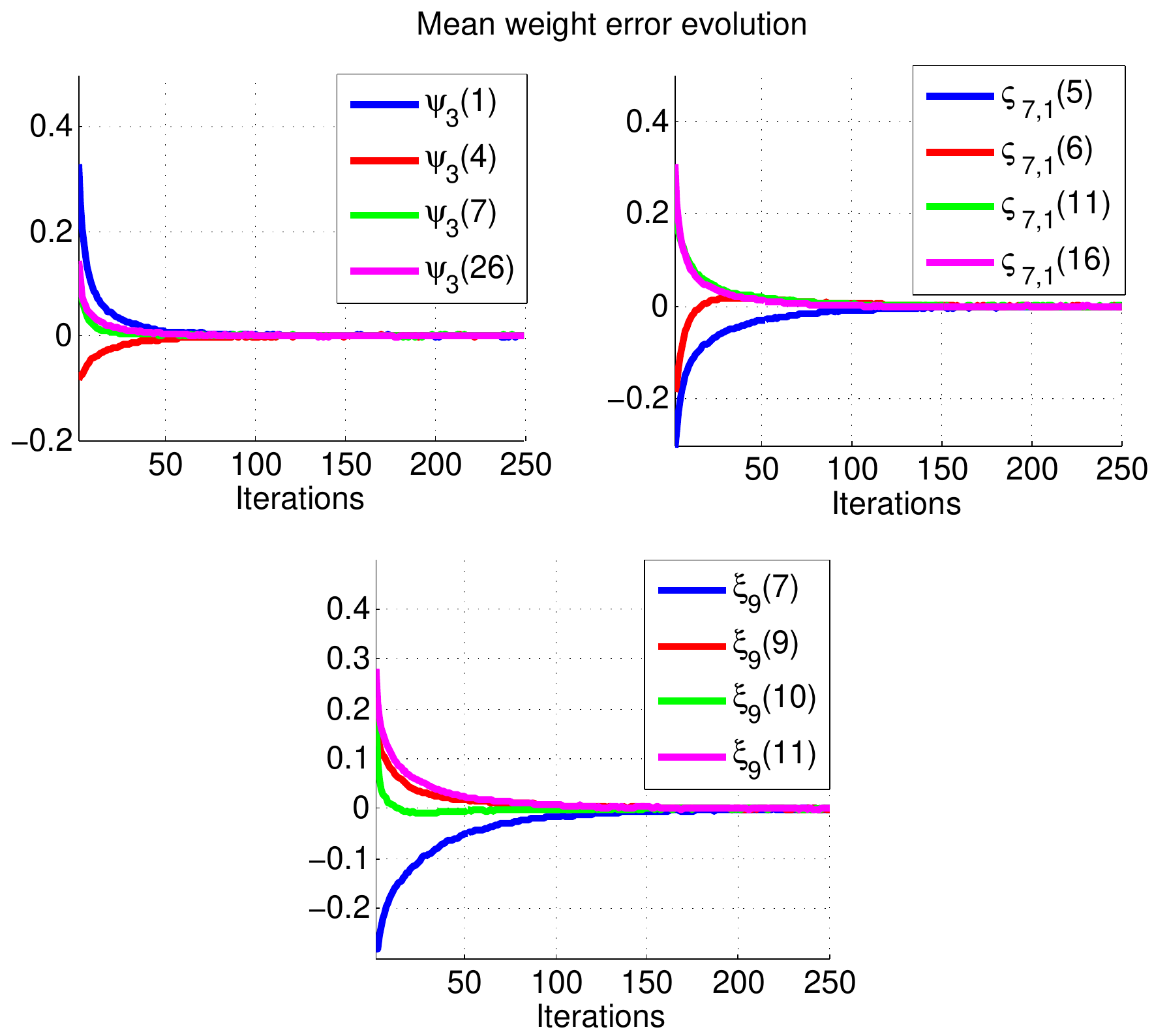,width=7.9cm}}
    \caption{The mean error trajectories of some vector coefficients related to the global (upper left), common (upper right) and local parameters (bottom) at randomly selected nodes.}
    \label{fig:ResultsMean}
\end{figure}

Finally, to illustrate the asymptotic unbiasedness of the proposed technique, in Fig.~\ref{fig:ResultsMean} we plot its mean weight behavior under the previously described setting. The figure indicates the mean weight evolution of some vector coefficients related to the global, common and local parameters at randomly selected nodes, whereas the optimal values from \eqref{eps:eq9} are indicated by the black lines. As expected by Theorem~\ref{teo:teo1}, D-NSPE LMS has estimated the optimum weight vectors without bias.

\section{Conclusions and future work}\label{sec:sec7}

We have addressed a novel NSPE problem where the estimation interests of the nodes consist of a set of local parameters, network-wide global parameters as well as common parameters to a subset of nodes. To do so, we have proposed two distributed adaptive schemes where a local LMS is run at each node in order to estimate each set of local parameters. Coupled among themselves and with all these local estimation processes, the parameters of global and common interests are estimated by LMS-based schemes implemented under a diffusion mode of cooperation. After obtaining conditions under which the proposed strategies are asymptotically unbiased, the mean-square steady-state performance has been evaluated. All the theoretical results have been validated through generic computer simulations. Moreover, the performance of the proposed algorithms have been illustrated in the context of cooperative spectrum sensing in Cognitive Radio networks.


\ifCLASSOPTIONcaptionsoff
  \newpage
\fi

%
%
%



%


\appendices

\section{}\label{ap:appendix1}

Here, we aim to specify the structure of the permutation matrix $\mathcal{P}$ in~\eqref{eps:eq_EWM_P1}. To this end, first note that there are $N$ blocks, each corresponding to a specific node, i.e., $\mathcal{P} = \text{col} \{ E_1, \ldots, E_N \},$ where the $k^{th}$ block $E_k$, of dimensions $M_k \times \breve{M}$, takes the following form
\begin{gather}\label{eps:eq_APP1_2_2}
\begin{split}
E_{k} = \mathrm{col} \{ & e_{f_g(k, 1)} , \ldots, e_{f_g(k, M_g)},  e_{f_c(k,\mathcal{I}_k(1),1)}, \\ 
&\ldots, e_{f_c(k,\mathcal{I}_k(1),M_c)}, \ldots, e_{f_c(k,\mathcal{I}_k(1),M_c)}, \\
&\ldots, e_{f_c(k,\mathcal{I}_k(|\mathcal{I}_k|),1)}, \ldots, e_{f_c(k,\mathcal{I}_k(|\mathcal{I}_k|),M_c)}, \\
&e_{f_l(k, 1)}, \ldots, e_{f_l(k, M_l)}   \}
\end{split}
\end{gather}
with the three counter functions, specifying the position of the unity in the basis vectors $e_{(\cdot)}$, defined by
\begin{gather}\label{eps:eq_APP1_3}
\begin{split}
f_g(k, c) = (k-1) \cdot M_g +c \, ,
\end{split}
\end{gather}
\begin{gather}\label{eps:eq_APP1_4}
\begin{split}
f_c(k, j, c) = N \cdot  M_g + \sum_{j'=1}^{j-1} |\mathcal{C}_{j'}| \cdot M_c +( |\mathcal{C}_{j,k}|-1) \cdot M_c + c
\end{split}
\end{gather}
and 
\begin{gather}\label{eps:eq_APP1_5}
\begin{split}
f_l(k, c) = N \cdot  M_g + \sum_{j=1}^{J} |\mathcal{C}_{j}| \cdot M_c + (k-1) \cdot M_l + c 
\end{split}
\end{gather}
with $\mathcal{C}_{j,k}$ given in~\eqref{eps:eq_def_cjk}. 

\bibliographystyle{IEEEtran}
\bibliography{IEEEabrv,./references}
\end{document}